\documentclass[12pt,aps,floatfix]{revtex4}
\usepackage{amsmath}
\usepackage{latexsym}
\usepackage{float}
\usepackage{amssymb}
\usepackage{graphicx}
\usepackage{epsfig}

\begin{document}
\title{Spherically averaged versus angle-dependent interactions in
quadrupolar fluids}

\author{B.\ M.\ Mognetti, M.\ Oettel, L.\ Yelash, P.\ Virnau, W.\ Paul
and K.\ Binder \\
Institut f\"ur Physik, Johannes Gutenberg--Universit\"at Mainz, \\
Staudinger Weg 7, D-55099 Mainz}

\begin{abstract}
Employing simplified models in computer simulation is
on the one hand often enforced by computer time limitations
but on the other hand it  offers insights into the molecular properties
 determining a given physical phenomenon. We employ this strategy
to the determination of the phase behaviour of quadrupolar fluids,
where we study the influence of omitting
angular degrees of freedom of molecules 
 via an effective spherically
symmetric  potential obtained from a perturbative expansion.
Comparing the liquid-vapor coexistence curve, vapor pressure at
coexistence, interfacial tension between the coexisting phases,
etc., as obtained from both the models with the full quadrupolar
interactions and the (approximate) isotropic interactions, we find
discrepancies in the critical region to be typically (such as in
the case of carbon dioxide)  of the order of 4\%.
However, when the Lennard-Jones parameters are rescaled such that
critical temperatures and critical densities of both models
coincide with the experimental results, almost perfect agreement
between the above-mentioned properties of both models is obtained.
This result justifies the use of isotropic quadrupolar potentials.
We present also a
detailed comparison of our simulations with a combined integral 
equation/density functional approach and show that the latter 
provides an accurate description except for the  vicinity 
of the critical point.
\end{abstract}
\maketitle

\section{INTRODUCTION}

The study of fluid phase equilibria by computer simulation methods
\cite{1,2,3,4,5} has become an extremely active field, since
accurate information on thermodynamic properties of simple and
complex fluids and their mixtures is of enormous importance for
a variety of applications \cite{6,7,8}. The problem of
understanding the phase behavior of such fluids is a
fundamental problem of statistical mechanics, too \cite{9,10}. While
such properties in principle can be found from experiments,
particularly for mixtures such data still are rather incomplete,
since a cumbersome study of a large space of control parameters
(temperature $T$, pressure $p$, and mole fraction(s) $x$ $(x_\alpha)$
in the case of binary (multicomponent) mixtures) needs to be made.

In simulations an economical use of computational
resources often dictates the use of models that are as simple as
possible. The standard approach for the simulation of fluids is to
apply classical Monte Carlo and Molecular Dynamics methods
\cite{1,2,11} that require effective potentials (usually of
pairwise type). Thus, both quantum effects associated with the
finite mass of the nuclei are ignored as well as the degrees of
freedom of the electrons (sometimes the latter are considered,
when the effective potentials are derived by ``ab initio'' quantum
chemistry methods, see e.g. \cite{12,13}, but sometimes these
potentials are postulated on purely empirical grounds
\cite{14,15}).

Now, even when the above approximations are accepted, there
the question remains whether an all-atom model is needed for the
description of intermolecular forces, or whether further degrees
of freedom may be eliminated. For example, consider carbon dioxide
(CO$_2$), which is an extremely important fluid due to its use as
supercritical solvent \cite{6,8}. Models used for the simulation
for CO$_2$ are truly abundant
\cite{14,16,17,18,19,20,21,22,23,24,25,26}; they include all-atom
models with either flexible or rigid intermolecular distances, as
well as models where a CO$_2$ molecule is reduced to a point
particle, with \cite{27} or even without \cite{28} a quadrupolar
moment. While the latter case, where the molecules are described
as simple point particles interacting with Lennard-Jones forces,
is computationally most efficient, it is also the least accurate.
Recently it was suggested \cite{27,29} that a significant gain in
accuracy with almost no loss in computational efficiency can be
obtained by using perturbation theory to construct an effective
isotropic quadrupolar potential \cite{SRN-74}. While very promising results for
a variety of molecular fluids, including carbon dioxide and
benzene, \cite{27} have been obtained, it  still needs to be
established to
what extent the isotropic quadrupolar potential actually
reproduces the physical effects of the actual angle-dependent
quadrupolar interactions.

In the present paper we fill this gap, using the case of CO$_2$ as
an archetypical example. Applying the same grand-canonical Monte
Carlo techniques in conjunction with successive umbrella sampling
\cite{30} and finite-size scaling analysis \cite{31,32} that were
used for the work applying the isotropic quadrupolar potential
\cite{27}, we obtain for the present model (which is described in
Sec.~2) the phase diagram in the temperature-density and
temperature-pressure planes, as well as the interfacial tension
(Sec.~3). In Sec.~4 we discuss the optimized choice of the
Lennard-Jones parameters, and show that the differences between
the models with the full and averaged quadrupolar interactions are
rather minor, when the critical temperatures and densities are
matched. Sec.~5 presents a comparison with integral 
equation/density
functional calculations and shows that the latter approach can
describe isotherms as well as the coexistence curve of the model
very accurately, but is not applicable very near to the critical
point. Sec.\ \ref{sec6} summarizes our conclusions.

\section{Model and Simulation Methods}\label{sec2}
\subsection{Models with Full and Averaged Quadrupolar Interaction}

We start from a system of uncharged point particles which have a
quadrupole moment $Q$ and interact also with Lennard-Jones (LJ)
forces,

\begin{equation}\label{eq1}
U_{ij}^{LJ} = 4 \epsilon\Bigg[\Big({\sigma \over r_{ij}}\Big)^{12}
-\Big({\sigma \over r_{ij}}\Big)^6\Bigg]\;,
\end{equation}

with $r_{ij}=|\vec{r}_i-\vec{r}_j|$ being the distance between
particles $i$ and $j$ at sites $\vec{r}_i$, $\vec{r}_j$, and the
range and strength of the LJ potential are denoted as $\sigma$ and
$\epsilon$, respectively.

The quadrupole-quadrupole interaction is

\begin{equation}\label{eq2}
U_{ij}^{QQ}= \frac {3Q^2}{4r_{ij}^5}f_{ij}^{QQ}\quad ,
\end{equation}

with \cite{33}

\begin{eqnarray}\label{eq3}
f_{ij}^{QQ}&=&1-5\cos ^2 \Theta _i -5 \cos ^2 \Theta _j + 17 \cos
^2 \Theta _i \cos^2\Theta_j+2 \sin ^2 \Theta _i  \sin ^2\Theta_j
\cos ^2(\Phi _i -\Phi _j) \nonumber  \\
&-& 16 \sin \Theta_i \cos \Theta_i \sin \Theta _j \cos\Theta _j
\cos (\Phi_i - \Phi _j)\;,
\end{eqnarray}

where ($\Theta _i,\Phi_i)$ are the polar angles characterizing the
orientation of the uniaxial molecule relative to the axis
connecting the sites $\vec{r}_i,\vec{r}_j$ of the two particles.

In order to speed up the Monte Carlo simulation, we wish to
introduce a cutoff $r_c$ such that the total potential is zero for
$r >r_c$. This needs to be done such that the total potential is
continuous for $r=r_c$, and the same condition should apply for
the equivalent isotropic quadrupolar potential, resulting from
treating the quadrupole-quadrupole interaction in second order
thermodynamic perturbation theory in the partition function
\cite{29,SRN-74}. As a result, we use the following potential (``F''
stands for ``full potential'')

\begin{equation}\label{eq4}
U_{ij}^F = \left\{
\begin{array}
{r@{\quad , \quad}l} 4 \epsilon
\Bigg[\Big(\frac{\sigma}{r_{ij}}\Big)^{12}-\Big(\frac{\sigma}{r_{ij}}
\Big)^6-\frac{3}{16}q_F\sqrt{\Big(\frac{\sigma}{r_{ij}}\Big)^{10}
-\Big(\frac{\sigma}{r_c}\Big)^{10}}
f_{ij}^{QQ} + S_0\Bigg]&
r \leq r_c \\
0 & r\geq r_c\;,
\end{array}
\right.
\end{equation}

where $S_0=(\sigma/r_c)^6-(\sigma/r_c)^{12}$. For $r_{ij} \ll r_c$
(and large enough $r_c$) Eq.~(\ref{eq4}) reduces to
Eqs.~(\ref{eq1})-(\ref{eq3}), noting the abbreviation

\begin{equation}\label{eq5}
q_F={Q^2\over (\epsilon \sigma ^5)}.
\end{equation}
Following previous investigations \cite{5,27,28}
we use $r_c=2\cdot$$^6\sqrt{2}\, \sigma$. Indeed in this work we 
will use some results  of Ref.\ \cite{27}, like the critical lines,
which depend on the choice of the cutoff. 
Differences in the thermodynamic properties arising from 
a different choice of the cut-off
are minor at least for simple Lennard-Jones potentials 
(and a suitable renormalization of
the LJ parameters) \cite{27}.

Note that the potential in Eqs.~(\ref{eq1})-(\ref{eq3}) is cut off
in Eq.~(\ref{eq4}) in such a way that the cutoff $r_c$ does not
depend on the angles ${\phi_i,\Theta_i}$. The corresponding
isotropic spherically averaged, potential is (``A'' stands for
``averaged potential'') \cite{29}

\begin{eqnarray}\label{eq6}
U_{ij}^{A}= \left \{ \begin{array}{r@{\quad , \quad}l} 4 \epsilon
_A \Bigg[\Big({\sigma _A\over r_{ij}}\Big)^{12}-\Big({\sigma_A\over
r_{ij}}\Big)^6 - \frac {7}{20}
q_A\Big({\sigma_A\over r_{ij}}\Big)^{10}+S_A \Bigg] & r \leq r_c  \\
0 & r \geq r_c , \end{array} \right.
\end{eqnarray}

where again the constant $S_A$ is chosen such that the potential
is continuous for $r=r_c$ \cite{27}. Note that for the potential,
Eq.~(\ref{eq6}), the forces at $r_c$ are discontinuous but do not
diverge there. This is a requirement if one wishes to estimate the
pressure from the virial theorem, when one does a simulation in
the $\mu$VT or NVT ensemble, respectively \cite{1,2}.

In Ref. \cite{27} we investigate extensively the use of potential
(\ref{eq6}) for modeling quadrupolar substances like carbon 
dioxide and benzene.
These results were compared with prior investigations of a simple LJ 
model \cite{28} without quadrupolar moment in which we only match 
critical temperature and density
with corresponding experimental values to obtain $\epsilon_A$ and $\sigma_A$.
(In principle, $\epsilon_A$ and $\sigma_A$ could also be matched at
 any point in the phase diagram,
which by definition would improve agreement around this point - however,
at the cost of imposing an inaccurate description of the critical region.)
The introduction of a spherically averaged quadrupolar 
moment improves agreement
with the experimental phase behaviour significantly, especially for
 carbon dioxide.
(Compare with Figs.\ \ref{fig6},  \ref{fig7} and \ref{fig8}).

The optimal choice of the parameters $\epsilon_A,\sigma _A$
and $q_A$ is somewhat subtle. A straightforward choice simply
requires that the two potentials are strictly equivalent for
temperatures $T \rightarrow \infty$, where the perturbation
expansion becomes exact. This would imply

\begin{equation}\label{eq7}
\epsilon_A = \epsilon, \qquad \sigma _A = \sigma, \qquad q_A =
{Q^4 \over k_B T \epsilon \sigma ^{10} }={\epsilon \over k_BT}
q^2_F .
\end{equation}

Note that for the potential $U_{ij}^A$ the parameter $q_A$ is not
a constant, but inversely proportional to temperature $T$.

However, the physically most interesting region of the system is
clearly not the regime $T \rightarrow \infty$, but rather the
vicinity of the critical temperature $T_c$. Thus, in our previous
work on carbon dioxide (CO$_2$) \cite{27} where Eq.~(\ref{eq6})
was used, we have chosen the parameters $\epsilon_A, \sigma _A$,
such that both $T_c$ and the critical density $\rho _c$ of the
model precisely coincide with their experimental counterparts
\cite{34} $T_{c,\exp}$ and $\rho_{c,\exp}$. Using also the
experimental value  of the quadrupole moment for CO$_2$,
$Q=4.3$  D$\mathrm{\AA}$, this implies \cite{27}

\begin{equation}\label{eq8}
\epsilon_A=3.491 \times 10^{-21}\, \mathrm{J},\quad \sigma_A=3.785
\, \mathrm{ \AA},\quad
q_{A,c}\equiv q_A(T_c) = 0.387,\quad q_F= 0.682 .
\end{equation}

In Sec.~3, we shall present numerical results for thermodynamic
properties of the full model, Eq.~(\ref{eq4}) and (\ref{eq5})
 with this choice of
parameters, Eq.~(\ref{eq8}), and compare them to the corresponding
results based upon Eq.~(\ref{eq6}) (some of the latter results
have been compared in \cite{27} to both experiment and simulations
of CO$_2$ using other models).

As we shall see in Sec.~3, in the critical region both models,
Eqs.~(\ref{eq4}) (\ref{eq5}) and Eq.~(\ref{eq6}) are no longer strictly
equivalent to each other, as expected since the accuracy of
perturbation theory deteriorates the lower the
temperature. Being interested in the critical region, it is more
natural to choose the parameters $\epsilon, \sigma$ and
$\epsilon_A, \sigma_A$ of both models such that $T_c,\rho_c$ of
both models match their experimental counterparts. This
requires necessarily a choice of $\epsilon$ and $\sigma$ different
from the choice implied by Eqs.~(\ref{eq7}), (\ref{eq8}), since the
latter choice would yield (choosing Lennard-Jones units $\varepsilon$,
$\sigma$  of Eq.\ \ref{eq8})
\begin{equation}\label{eq9}
T_c^{*(F)} = 1.167, \quad \rho_c^{*(F)} = 0.340,
\end{equation}
\begin{equation}\label{eq10}
T_c^{*(A)}  = 1.203, \quad \rho_c^{*(A)} = 0.347,
\end{equation}
as shall be discussed in more detail in Sec.~3.

Since the relation between $q_F$ and $Q$ (Eq.~\ref{eq5}) or
$q_A$ and $Q$ (Eq.~\ref{eq7}) depends on $\epsilon$ and
$\sigma$ as well, finding the choice of the latter parameters
which yields $T_c= T_{c,\exp}$ and $\rho_c= \rho_{c,\exp}$ is a
self consistency problem \cite{27}. In principle, one needs to
record the functions $T^*_c(q_F)/T^*_c(q_F=0)$ and
$\rho^*_c(q_F)/\rho^*_c(q_F=0)$, similarly as described in \cite{27}.
However, since the differences between the results of
Eqs.~(\ref{eq9}) and (\ref{eq10}) are rather small, it is a very
good approximation to simply keep the value of $q_F$ as found in
Eq.~(\ref{eq8}) and just recompute the appropriate values of
$\epsilon$ and $\sigma$, which we shall denote as $\epsilon_F$ and
$\sigma_F$, in order to distinguish them from the choice of
Eqs.~(\ref{eq7},\ref{eq8}). This procedure immediately yields

\begin{equation}\label{eq11}
\epsilon_F = 3.598 \times 10^{-21}\, \mathrm{J}\;,\quad \quad \sigma_F = 3.760
\,\mathrm{\AA}\quad .
\end{equation}

A good test of possible errors introduced by this approximation is
provided by using Eq.~(\ref{eq5}) together with Eq.~(\ref{eq11})
to check the precise value of the physical quadrupole moment
strength $Q$ this corresponds to. This yields $Q_{\textrm{new}} =
4.292$  D$\mathrm{\AA}$ instead of the value used in \cite{27}, $Q=4.300$
 D$\mathrm{\AA}$ (note that the actual quadrupole moment of CO$_2$ is
negative, but since only the square of $Q$ actually
matters, cf.\ Eq.~(\ref{eq2}), we suppress the sign of $Q$
throughout). If one uses this slightly modified value
$Q_{\textrm{new}}$ instead of $Q$ in Eq.~(\ref{eq7}) together with
the master-curves $T^*_c(q_A(T_c))/T^*_c(0)$ and
$\rho^*_c(q_A(T_c))/\rho^*_c(0)$ calculated in \cite{27}, instead of
Eq.~(\ref{eq8}) slightly revised estimates of $\epsilon _A$ and
$\sigma_A$ would result

\begin{equation}\label{eq12}
\epsilon_A = 3.494 \times 10^{-21}\, \mathrm{J} \quad , \quad \sigma _A =
3.784 \, \mathrm{\AA}, \quad q_A(T_c)= 0.385\quad .
\end{equation}

But in view of the large error  with which the actual quadrupole
moment strength of CO$_2$ is known \cite{34}, $Q=4.3 \pm 0.2$
 D$\mathrm{\AA}$, Eq.~(\ref{eq12}) is as good as a representation of reality
as the choice of \cite{27} (as quoted in Eq.~(\ref{eq8})) has
been.

In Sec.~4, we shall compare the result of the model with the full
quadrupolar interaction Eq.~(\ref{eq4}), using Eq.~(\ref{eq11}) as
LJ parameters, with the averaged interaction, Eq.~(\ref{eq6}),
using Eq.~(\ref{eq12}) as choice for the LJ parameters, since then
all physical parameters $(T_c, \rho_c, Q)$ that coincide with
their experimental counterparts, have precisely the same values.

\subsection{Simulation Methods and Tools for the Analysis of the Simulation Data}

In this section, we summarize our procedures for carrying out the
simulations and their analysis only very briefly, since more
detailed descriptions for similar models can be found in the
literature \cite{5,27,28}.

The estimation of vapor-liquid coexistence curves and critical
parameters is done in the grand canonical ($\mu$VT) ensemble,
varying the chemical potential $\mu$ and recording the density
distribution $P_L(\rho)$, and analyzing carefully the dependence
on the linear dimension $L$ of the simulation box (as usual, we
take $V=L^3$ at the critical point, while deep into the coexistence 
region we use an elongated box $V=2\cdot L^3$. For both geometries
 periodic boundary conditions are applied). For $T
<T_c$, the value of $\mu_{\textrm{coex}}(T)$ where phase
coexistence between vapor and liquid occurs is found from the
``equal weight rule'' \cite{30,31,32,35}. For an accurate sampling
of $P_L(\rho)$ including the densities inside the two phase
coexistence region that also need to be studied for an accurate
estimation of the weights of the vapor and liquid phases,
successive umbrella sampling methods \cite{28,30} are used, as
well as re-weighting procedures \cite{30,31,32,36}. Note that the
presence of the orientational degrees of freedom in
Eqs.~(\ref{eq3},\ref{eq4}) does not constitute any principal
difficulty here. The acceptance rate for the insertion of
particles with a randomly chosen  orientation  is of the same order 
as for the isotropic potential,
Eq.~(\ref{eq6}), where this degree of freedom has been eliminated.
This fact is understood easily, since the strength of the
quadrupolar interaction, Eq.~(\ref{eq2}), is distinctly smaller
than the strength of the Lennard-Jones interaction,
Eq.~(\ref{eq1}), for the present choice of $q_F$. However, the
time required to compute the energy change caused by such a
particle insertion or deletion is about an order of magnitude
larger when Eq.~(\ref{eq4}) rather than Eq.~(\ref{eq6}) is used,
due to the complicated angular dependence of the
quadrupole-quadrupole interaction (Eq.~\ref{eq3}).

\begin{figure}\caption{\label{fig1} }
\includegraphics[angle=-0,scale=0.7]{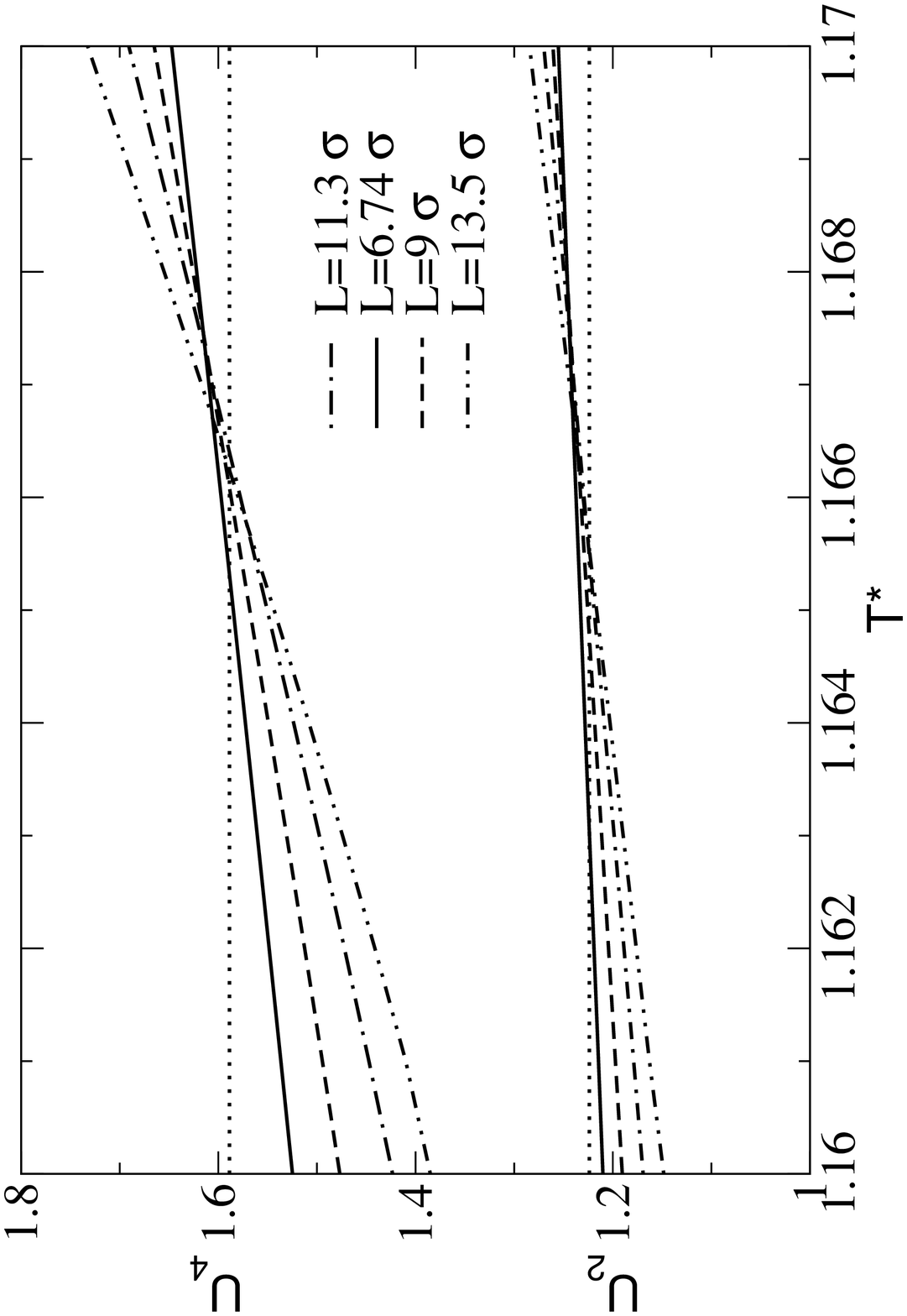}
\end{figure}

Nevertheless it is still feasible for this model, Eq.~(\ref{eq4}),
to obtain sufficiently accurate information on $P_L(\rho)$ for a
variety of temperatures $T$ and lattice linear dimensions $L$,
following a path along the coexistence curve
$\mu=\mu_{\textrm{coex}}(T)$ in the $(\mu,T)$ plane, and its
continuation for $T>T_c$ (there the path is defined by the
condition that the derivative $(\partial \rho/\partial \mu)_T=
L^3(\langle \rho^2\rangle _{T,\mu}-\langle \rho \rangle
_{T,\mu}^2)$ is maximal). Fig.~\ref{fig1} shows, as an example,
second and fourth order cumulants $U_2$ and $U_4$ along such a
path as a function of temperature. These cumulants are defined by

\begin{equation}\label{eq13}
U_2=\langle M^2 \rangle /\langle |M|\rangle ^2, \qquad U_4= \langle
M ^4 \rangle /\langle M^2\rangle ^2, \qquad M = \rho-\langle \rho
\rangle ,
\end{equation}
\\
where now we have omitted the subscripts $(T,\mu)$ from the
averages $\langle \ldots \rangle$. As is well known
\cite{4,5,11,31,32}, accurate estimates for $T_c$ can
be obtained from the common intersection point of either
$U_2(L,T)$ or $U_4(L,T)$ for different $L$. The justification of
this simple recipe follows from the theory of finite size scaling
\cite{37,38,39}. Note also that the ordinate values $U^*_2,U^*_4$
of these common crossing points should be universal for all
systems belonging to the Ising model universality class, to which
both models Eq.~(\ref{eq4}) and (\ref{eq6}) should belong, and are
known with very good accuracy \cite{4,40}.

From Fig.~\ref{fig1} it is evident that this intersection property
does not work out perfectly well, in particular, the curve for
$L=6.74\, \sigma$ is somewhat off. However, finite size scaling should
become exact only in the limit where both $L \rightarrow \infty$
and $T\rightarrow T_c$, while otherwise corrections come into
play. Systematic improvements (taking the so-called ``field
mixing'' \cite{4} and ``pressure mixing'' effects \cite{41} into
account) are possible, but are not considered to be necessary
here, since the relative accuracy of our estimate for $T_c$
extracted from Fig.~\ref{fig1} is clearly not worse than
3$\cdot$10$^{-3}$, and this suffices amply for our purposes. 
A recent comparative study of different finite size scaling
based approaches for the study of critical point estimation of
Lennard Jones models \cite{41p} is in full agreement with this 
conclusion.
Note that
the accuracy of the data in Fig.~\ref{fig1} is comparable to data
taken for a pure LJ model \cite{28} or for Eq.~(\ref{eq6})
\cite{27}, respectively (we estimate the relative accuracy of the
curves in Fig.\ \ref{fig1} to be of the order of 0.5\% or better).
We have  used 7$\cdot$10$^6$, 3$\cdot$10$^6$, 6$\cdot$10$^6$ and 
9$\cdot$10$^6$   Monte Carlo steps (respectively for the $L/\sigma=6.74$, 9,
11.3 and 13.5 system) for each simulation point ($T^*$,$\Delta\mu_\mathrm{coex}
(T^*)$), for which the data for $P_L(\rho)$ were sampled, and applied
histogram extrapolation methods (see \cite{27,28,32} for details) to
obtain the smooth curves drawn in Fig.\ \ref{fig1}. In every step  
100 attempts to insert or delete
particle plus local moves are done.

For $T<T_c$ the densities $\rho_{\textrm{coex}}^{(1)},
\rho_{\textrm{coex}}^{(2)}$ of the two coexisting vapor and liquid
phases can be simply read off from the peak positions of
$P_L(\rho)$, and from the density minimum in between the peaks the
interfacial tension $\gamma (T)$ can be estimated, using the
relation

\begin{equation}\label{eq14}
\gamma (T)/k_BT=0.5 L^{-2}\ln
[P_L(\rho_{\textrm{coex}}^{(1,2)})/P_L(\rho_d)],\qquad \qquad  L
\rightarrow \infty
\end{equation}

where $\rho_d=(\rho_{\textrm{coex}}^{(1)}+
\rho_{\textrm{coex}}^{(2)})/2$ denotes the density of the
``rectilinear diameter''. All these methods work well for fluid
models with simple isotropic potentials \cite{27,30} and their
extension to the present model (Eqs.~\ref{eq3}, \ref{eq4}) is
fairly straightforward.

For a comparison between the two models defined by
Eqs.~(\ref{eq4}) and (\ref{eq6}) outside
of the critical region it is also of interest to apply NVT and NpT
ensembles. Then no particle insertions or deletions occur, but
rather a particle is selected at random and a move is attempted
where one puts it in a randomly chosen position inside a sphere of
radius $\delta r$ around its old position. Simultaneously the
orientation of its molecular axis is chosen inside a cone of angle
$\delta \Theta$ around its old orientation \cite{1,2}. The choices
of $\delta r$ and $\delta \Theta$ were adjusted to have an
acceptance rate of 40\% for such moves.

\section{Direct Comparison Between Results for the Full and Corresponding Averaged Quadrupolar Potential}

As noted in Sec.~2.1, Eq.~(\ref{eq6}) results from
Eqs.~(\ref{eq3},\ref{eq4}) when one carries out a second order
thermodynamic perturbation theory and interprets the result as
being due to an average potential \cite{29}. Since thermodynamic
perturbation theory is basically a high temperature expansion in
powers of $1/T$, it is a matter of concern how accurate such a
procedure really is in the temperature region around criticality
and below.

As a first test, we have carried out a NVT simulation using the averaged
model, Eq.~(\ref{eq6}), at a density that is much larger than the
critical density, namely $\rho^*= 0.544$, and we have recorded the
corresponding pressure $p^*(T)$ from the virial formula
(Fig.~\ref{fig2}a). This pressure then was used as an input for a
NpT simulation of the full model, Eqs.~(\ref{eq3},\ref{eq4}).
Fig.~\ref{fig2}(b) shows the corresponding comparison: one sees
that for large $T^*$ (i.e., $T^* \geq 2.5$), the data obtained
from the full potential indeed converge against the density
$\rho^*$ that was chosen, while for $T^*\leq 1.5$ there are rather
pronounced deviations. Of course, if an NpT simulation is carried
out using Eq.~(\ref{eq6}), the chosen density $\rho^*=0.544$ is
reproduced over the entire temperature interval shown in
Fig.~\ref{fig2}(b) with negligibly small errors, since with the
chosen volume ($L/\sigma_A=10.3$) systematic discrepancies between the
different ensembles of statistical mechanics are completely
negligible (although such discrepancies will occur in the
two-phase coexistence region or near the critical point). The
deviations seen in Fig.~\ref{fig2}(b) simply represent the higher
order terms of the $1/T$ expansion, by which the averaged and full
potentials differ. Similar discrepancies between the full and
averaged potential were also seen on the vapor side of the
coexistence curve (not shown here).

As a result, coexistence densities in the $(T^*, \rho^*)$ plane
are rather different as well for the two models, Fig.~\ref{fig3},
as expected from the differences noted in
Eqs.~(\ref{eq9},\ref{eq10}), and a similar discrepancy occurs
between the predictions for the pressure along the coexistence
curve (Fig.~\ref{fig4}) and the interfacial tension
(Fig.~\ref{fig5}).

\section{Comparison between Results for the Full and Averaged Quadrupolar Potentials with Optimized Parameters}

Now we present a comparison between the full quadrupolar plus
Lennard-Jones potential (Eqs.\ \ref{eq4}, \ref{eq5}) and the spherically
averaged one (Eq.\ \ref{eq6}) choosing the parameters as given in
Eq.\ (\ref{eq11}) for the full potential and in Eq.\ (\ref{eq12}) for
the averaged one, for which critical temperatures $T_c$ and critical 
densities $\rho_c$ coincide with their experimental counterparts
in both cases.

Fig.\ \ref{fig6} shows that along the  vapor branch of the
vapor-liquid coexistence curve the averaged potential 
slightly underestimates the experimental vapor densities, 
while the full potential slightly overestimate them. However,
these deviations are of the same order in both cases, and hardly 
visible (on the scale of Fig.\ \ref{fig6}) anyway. 
Recalling also the fact that the coexistence curves (and other
data extracted from the simulation) still may suffer from 
systematic effects (residual finite size effects) and statistical
errors, see Sec.\ \ref{sec2}, of the order of up to 0.5\%, one
should not pay too-much attention to the residual differences. 
We conclude that
both models describe the vapor branch of the coexistence curve
equally well, over the studied range of temperatures (which extends from
$T_c$ down to about $T=250$ K).

However, for the liquid branch of the coexistence curve the model
with the averaged interaction performs distinctly better. Of
course, there is no physical reason known to us why this should 
be the case. We believe that this more accurate description of the
isotropically averaged model is only due
to a fortunate compensation of errors.

With respect to the vapor pressure at phase coexistence (Fig.\
\ref{fig7}), we see, however, that at low temperature (250 K
 $\le T \le$ 280 K) the model with the full quadrupolar 
interaction performs slightly better than the isotropically averaged model.
Near the critical point, however, the isotropic quadrupolar 
interaction performs slightly better, since it predicts the
critical pressure a bit more accurately. Thus, we conclude
that the vapor pressure at coexistence is predicted by both
models about equally well.

Fig.\ \ref{fig8} finally compares both model predictions
with the data \cite{34} for the surface tension between
both phases. In this case there is a clear preference 
for the model with the isotropically averaged quadrupolar 
interaction. Taking the results from Figs.\ 
\ref{fig6}-\ref{fig7} together, we conclude that for a
description of phase coexistence the model Eq.\ (\ref{eq6})
is clearly the better ''effective'' model. Also 
for a supercritical isobar (Fig.\ \ref{fig10} presents an 
example) the model Eqs.\ (\ref{eq4},\ref{eq5},\ref{eq11})
does not have an advantage. The comparisons presented in this 
section thus fully justify the use of Eq.\ (\ref{eq6})
for practical purposes.

An additional interesting test now concerns the
temperature dependence of the density that results when
we compare NVT simulations for the averaged potential with
NpT simulations for the full potential (similarly to
what was done in Fig.\ \ref{fig2}), choosing the parameters
of Eq.\ (\ref{eq11}) and (\ref{eq12}). 
We have also included a comparison with two analytical 
approaches, namely an integral equation/density functional
theory (IE/DF), see the following section, and perturbation theory
combined with mean spherical approximation (MSA), see Ref.\ \cite{27}
for a description of this method in the current context. Both
approaches agree with our results very well. 
Fig.\ \ref{fig9}(a) is the
counterpart of Fig.\ \ref{fig2}(a); again it is seen that
the pressure at $\rho=0.733$ g/cm$^3$ for the full model
is in very good agreement with the corresponding experimental
data and averaged potential. However in this case the full model
is superior with respect to the averaged model. Fig.\ \ref{fig9}(b)
shows that indeed the NpT results for the full potential
now converge rapidly to a somewhat higher density (near
$\rho\approx 0.75$ g/cm$^3$) as the temperature is raised
from the critical region to higher temperatures. Of course,
as expected, it is not possible to fit the critical region
(as done in Fig.\ \ref{fig6}-\ref{fig9}) and the high temperature
region (as done in Figs.\ref{fig2}-\ref{fig5}) simultaneously.

\section{Integral Equations with Reference Functionals}\label{IE/DF}

In this section we summarize a combined integral equation/density 
functional method to calculate
equations of state. A novel approach to avoid unphysical no--solution 
domains near the critical
point is outlined and   
data for the equation of state of the averaged model 
defined by Eq.~(\ref{eq6})
are compared with the simulation results.

The pair correlation function $h(r)$ and the direct correlation function
(of second order)
$c( r)$, which  contain all thermodynamic information of a given homogeneous model system
of density $\rho_0$ and interacting with an isotropic pair 
potential $U_{ij}(r)$, 
fulfill the following general relations \cite{Mor60,Ste64,Per64}:
\begin{eqnarray}
  \label{eq:oz}
  h( r) - c( r) &=& \rho_0 \int d^3 r'\; h(|{\bf r}-{\bf r}'|)\, c( r') \;, \\
  \label{eq:cl}
  \ln[1+h(r)] - \beta U_{ij}( r) &=&  h(r) - c(r) - b( r)\;.
\end{eqnarray}
The first relation is known as the Ornstein--Zernike equation,
 while the second is the general closure
relation in terms of a yet unknown bridge function $b(r)$. 
For a solution, it is necessary to specify
the bridge function $b$ in terms of $h$ and $c$. 
Thermodynamics is obtained either through the virial 
route \cite{Han06}, giving the pressure $p$,
\begin{equation}
  \label{eq:virial}
  \frac{\beta p}{\rho_0} = 1 - \frac{2}{3}\, \pi\, \rho_0\,  \beta \int_0^\infty
    dr\, r^3 \,(1+h(r)) \frac{d U_{ij}(r)}{dr} \; ,
\end{equation}
 or through the compressibility route \cite{Han06}
\begin{equation}
 \label{eq:com}
  \beta \frac{\partial p}{\partial \rho_0} = 1-4\pi\rho_0 \int_0^\infty dr\, r^2\,c(r) \;.
\end{equation}
Both routes are not necessarily identical
for a given approximation for the bridge function. 

For one--component systems, advanced methods exist which yield good agreement with simulations
for the pressure in the whole $\rho-T$ plane and for the whole coexistence
curve, e.g.\ 
the SCOZA (self--consistent Ornstein--Zernike approximation) \cite{Pin02} or the HRT (hierarchical
reference theory) \cite{Par95}. A drawback of these methods is their ``non--locality" in the 
$\rho-T$ plane, i.e. for a SCOZA solution a partial differential equation has to be solved on
this plane and for a 
HRT solution a renormalization flow equation has to be solved
 on a specified isotherm. 

Therefore we treat our problem within a formalism which 
is close to the reference hypernetted-chain (RHNC) method.  
In its original version \cite{Lad83} (developed for repulsive core fluids), 
the bridge function $b$ was taken from a reference hard sphere 
system with suitably optimized
reference packing fraction (or hard sphere diameter). 
Since the RHNC closure equation (\ref{eq:cl}) can be derived
from an approximate bulk free energy functional, a closed expression 
for the chemical potential
$\mu$ also exists. Near the critical point, however, 
there exists a no--solution domain (extending into
the supercritical region $T>T_c$) where
no real solution to the RHNC closure can be found. As shown below, this problem can be eliminated 
by a method (FHNC) where
a generating {\em functional} for the bridge function is adopted 
from a suitable reference system.

\subsection{Bridge functions from a reference functional}

The FHNC generalization of RHNC within the framework of density functional theory was proposed 
in Ref.~\cite{Ros93}. 
By making a Taylor expansion of the free energy functional 
around bulk densities the general closure equation is derived and
the bridge function in the bulk is determined via a density functional 
for a suitably chosen
reference system of hard spheres. To be more explicit, 
the excess free energy functional
\begin{equation}
 {\cal F}^{\rm ex}[\rho({\bf r})] =  {\cal F}^{\rm HNC}[\rho({\bf r})] +   {\cal F}^{\rm B}[\rho({\bf r})]
\end{equation}
is split into a part (${\cal F}^{\rm HNC}$) which generates the
hypernetted-chain (HNC) closure ($b=0$) and a remainder
(${\cal F}^{\rm B}$) which generates a non-zero bridge function. 
The HNC part
${\cal F}^{\rm HNC}$ is a Taylor expansion of
the excess free energy up to second order in the deviations 
from the bulk density $\rho_0$, 
$\Delta\rho({\bf r}) = \rho({\bf r})-\rho_0$:   
\begin{eqnarray}
  \label{eq:fhnc}
  {\cal F}^{\rm HNC} = F(\rho_0) + \mu^{\rm ex}\int d^3r \,\Delta\rho({\bf r}) -
    \frac{1}{2\beta} \int d^3r \int d^3r' c(|{\bf r}-{\bf r}'|) \,\Delta\rho({\bf r})\,\Delta\rho({\bf r'}) \;.
\end{eqnarray}
There the defining relations for the excess chemical potential $\mu^{\rm ex}(\rho_0)$ and the
direct correlation function $c$ have been used: 
\begin{eqnarray}
  \left.\frac{\delta {\cal F}^{\rm ex}}{\delta \rho({\bf r})}\right|_{\rho({\bf r})=\rho_0} = 
   \mu^{\rm ex}(\rho_0), \qquad
  \beta\left.\frac{\delta^2 {\cal F}^{\rm ex}}{\delta \rho({\bf r}) \delta \rho({\bf r'})}\right|
  _{\rho({\bf r})=\rho({\bf r'})=\rho_0} = - c(|{\bf r}-{\bf r}'|;\rho_0)\;.
\end{eqnarray}
The general closure equation follows by employing the test particle method: the grand potential 
\begin{eqnarray}
  \Omega[\rho({\bf r})] & = & {\cal F}^{\rm id}[\rho({\bf r})] + {\cal F}^{\rm ex}[\rho({\bf r})]
 - \int d^3r\,(\mu - V({\bf r}))\rho({\bf r}) \\
    && \left(\beta {\cal F}^{\rm id}[\rho({\bf r})] = 
   \int d^3r\, \rho({\bf r})(\ln[ \Lambda^3\rho({\bf r})] -1) \; \right)
  \nonumber
\end{eqnarray}
is minimized in the presence of a fixed test particle of the same 
species which exerts
the external potential $V({\bf r})\equiv U_{ij}(r)$ onto the 
fluid particles \cite{Ros93,Oet05}
($\Lambda$ is the thermal de-Broglie wavelength). The precise
form of the closure equation  (Eq.~\ref{eq:cl}), is recovered upon 
the following identifications:
\begin{eqnarray}
  h(r) = {\Delta\rho(r)\over \rho_0}, \qquad b(r) = \beta
\left. \frac{\delta {\cal F}^{\rm B}}
  {\delta \rho({\bf r})}\right|_{\rho({\bf r})= \rho_0(h(r)+1)} \;.
\end{eqnarray}
In general, the excess free energy functional beyond second order,
the bridge functional  ${\cal F}^{\rm B}$, is not known.
Therefore the key step of the present method is to 
approximate ${\cal F}^{\rm B}$ 
by a density functional for a reference system in the following
manner:
\begin{eqnarray}
 \label{eq:fbref}
 {\cal F}^{\rm B}[ \rho] \approx {\cal F}^{\rm B,ref}[ \rho] &=&
    {\cal F}^{\rm ref}[ \rho]-
          {\cal F}^{\rm HNC,ref}[ \rho]\;, 
\end{eqnarray}
where the second order HNC contribution is defined as in Eq.~(\ref{eq:fhnc})
with the replacements $F(\rho_0) \to F^{\rm ref}(\rho_0)$, $\mu^{\rm ex} \to \mu^{\rm ex,ref}$ and $c \to c^{\rm ref}$. For fluids with repulsive cores, the reference functionals of choice are hard--sphere
functionals which are known with high accuracy (such as e.g. the ones in Refs.~\cite{Ros89,Rot02}).
In such a manner the system of equations (\ref{eq:oz}) and (\ref{eq:cl}) is closed and amenable to
numerical treatment. According to
Ref.~\cite{Oet05} the optimal reference hard sphere packing fraction 
$\eta^{\rm ref} = (\pi/6)\,\rho_0\,d_{\rm ref}^3$ is determined 
through the (local) condition
\begin{equation}
 \label{eq:crit2}
  \frac{\partial}{\partial d_{\rm ref}} \left( {\cal F}^{\rm B,ref}[ \rho_0\: (h(r)+1);d_{\rm ref}]
   - {\cal F}^{\rm B,ref}[ \rho_0\: (h^{\rm ref}(r)+1);d_{\rm ref}] \right)
    \stackrel{!}{=} 0 \;,
\end{equation}
which corresponds to extremizing the free energy difference between 
the fluid
and the reference system with respect to the 
reference hard sphere diameter $d_{\rm ref}$.
 The 
chemical potential of the fluid can also be expressed as
 a functional of $h$ locally in the $\rho-T$ plane
\cite{Oet05}.
Thus a coexistence curve for a given fluid can be determined straightforwardly by the equality of
$p$ and $\mu$ on the fluid and the gas side, respectively, and no thermodynamic integrations are
necessary.

\subsection{The critical region}

\label{sec:crit}

Similarly to RHNC and HNC, the FHNC method outlined above, together with the optimization
criterion for $\eta^{\rm ref}$, Eq.~(\ref{eq:crit2}), exhibits
a no--solution domain in the $\rho-T$ plane which stretches into the supercritical region ($T>T_c$).
This can be attributed to a failure of the optimization criterion in the critical region which 
assigns a wrong long--range behavior to the direct correlation function $c$. Consider
the asymptotic expansion of the closure, Eq.~(\ref{eq:cl}), where $h$, $c$ and $b$ are small:
\begin{equation}
 \label{eq:cl_asy}
    - \frac{h^2(r)}{2} + \frac{h^3(r)}{3} + \dots = - c(r) - b(r) \qquad (r\to\infty)\;.
\end{equation} 
(Here we assume that the potential $U_{ij}(r)$ is cut off.) In HNC (RHNC) the bridge function
$b$ is zero (short--ranged), therefore we find to leading order $ c(r) \approx h^2(r)/2$.  
However this is inconsistent with the critical behavior of the correlation functions. In three
dimensions, this critical behaviour is approximately described by the Ornstein--Zernike form
\cite{Stanleybook}
\begin{equation}
 \label{eq:h_oz}
   h^{\rm OZ}(r)  \to \frac{\exp(-r/\xi)}{r} 
\end{equation}
Through the Ornstein--Zernike equation (\ref{eq:oz}) it follows that in this limit $c^{\rm OZ}(r)$
stays short ranged in the sense that its Fourier transform 
$\tilde c^{\rm OZ}(q)=c_0+c_2\, q^2+ \dots$ 
permits an expansion around $q=0$.
In Eq.~(\ref{eq:h_oz}), $\xi$ is the correlation length which goes to infinity upon 
approaching the critical point and is related to $\tilde c^{\rm OZ}$ through
$\xi^2= -\rho_0 c_2/(1-\rho_0 c_0)$.
 Assuming $h\to h^{\rm OZ}$, the asymptotic HNC (RHNC) closure
$c\approx h^2/2$ is in conflict with the requirement of $c$ staying short--ranged upon
approaching the critical point, i.e. it cannot be a solution of the Ornstein--Zernike equation.
On the other hand, within FHNC the bridge function $b$ itself depends on $h$
as
\begin{equation}
  b(r) \approx (\eta^{\rm ref})^2 \frac{\beta}{2} \frac{\partial^2 \mu^{\rm ex, ref}}{\partial (\eta^{\rm ref})^2}\;h^2(r)
   + {\cal O}(h^3)
  \qquad (r\to\infty) \;.
\end{equation}      
and the asymptotic closure, Eq.~(\ref{eq:cl_asy}) reads
\begin{equation}
  c(r)  = \frac{1}{2}\left(1-(\eta^{\rm ref})^2 \beta\frac{\partial^2 \mu^{\rm ex, ref}}{\partial (\eta^{\rm ref})^2} \right)
   \;h^2(r) + {\cal O}(h^3)\;.
\end{equation}
Thus we see that upon requiring $\beta(\eta^{\rm ref})^2 \frac{\partial^2 \mu^{\rm ex, ref}}{\partial (\eta^{\rm ref})^2}
\stackrel{!}{=} 1$ the closure is consistent with $h \to h^{\rm OZ}$ and $c$ staying short--ranged.
This condition is fulfilled for $\eta^{\rm ref} \approx 0.13$ and in the critical region, 
this condition on the reference
system packing fraction replaces Eq.~(\ref{eq:crit2}). Incidentally, this condition is
consistent with the intuition that the reference hard sphere diameter $d_{\rm ref}$ is
roughly equal to the Lennard--Jones diameter $\sigma$ for densities close to the critical density. 
We checked for various supercritical isotherms that
the modified optimization criterion indeed removes the no--solution domains. (For a different
approach to this problem within FHNC, see Ref.~\cite{Amo06}.)

\subsection{Numerical results}

Numerical data for the coexistence curve (virial route) of the 
averaged model (Eq.~\ref{eq6}) 
are given in Fig.~\ref{fig3} ($q_A=0.387$) and Fig.~\ref{fig6} 
($q_A=0.385$). 
The overall agreement with the
simulation results is good, except for temperatures within 5\% of 
the critical
temperature $T_c$ where a noticeable shift of the
 coexisting gas densities towards higher values
can be observed. This is also the reason why the pressure 
on the coexistence curve is somewhat larger
than the pressure determined by the simulations (Fig.~\ref{fig7}). 
Analyzing the behavior of the solutions
in the critical region more closely, we find that there (as many 
other integral equation approaches) 
the FHNC method suffers from the inconsistency 
between the virial and the compressibility route to the equation of state. 
Indeed, the compressibility route
(for $q_A=0.387$)
gives a critical temperature of $T_c^{\rm com} \approx 1.14$
 and critical density
$\rho_c^* \approx 0.36$ with a behavior of the correlation
function $h$ consistent with the Ornstein--Zernike form, Eq.~(\ref{eq:h_oz}),
confirming the applicability of the reasoning in the previous subsection. 
Therefore, the virial route coexistence data for temperatures 
above $T_c^{\rm com}$ are not reliable.

We remark that the FHNC method is not particularly designed to
 capture the correct critical behavior.
With respect to the prediction of the coexistence curve and coexistence 
pressure only, it is not particularly
superior in accuracy to the (simpler and faster) perturbative
Mean Spherical Approximation (MSA)
whose results have been discussed in \cite{27} 
(for a detail description of the Equation of State  used in  \cite{27},
we refer to  \cite{MMVB00,MVMB02}).
Clearly, in this respect renormalization--group based methods
 such as HRT perform much better. 
However, supercritical properties of CO$_2$ are reproduced quite
 accurately as the comparison of the $p-\rho$ isotherms ($T=316.36$ K) between the experimental data
for CO$_2$, FHNC and MSA shows (Fig.~\ref{fig11}). While the experimental data and the 
FHNC results are almost on top of each other, the perturbative 
MSA results exhibit a van--der--Waals
loop due to the underlying mean--field approximation which results in a too large 
$T_c$ \cite{27}. 
Additionally we observe
 very good agreement with simulations for the
supercritical isochore $\rho=0.733$ g/cm$^3$ (Fig.~\ref{fig9}) and the 
supercritical isobars p=100 bar and 200 bar (Fig.~\ref{fig10}).

A problem of the FHNC approach seems to be the accurate prediction
of surface tensions. Although the technique can be extended to
compute this quantity, the results are much less satisfactory, since
the simulation results are about 40\% lower than the FHNC results, in the
temperature region around T=270 K where the coexistence curve is
predicted rather satisfactorily by FHNC (Fig.\ \ref{fig6}). A similar
problem was noted in a recent comparison of Monte Carlo and density functional
theory results for phase separation in colloid-polymer mixtures
\cite{50p}.

In concluding this section, we find that the presented FHNC method allows a fast and 
precise determination of the equation of state
except for the vicinity of the critical point. Within FHNC, the pressure $p$ and the chemical
potential $\mu$ are obtained through local relations in the $T-\rho$ plane.
It appears as an advantage that FHNC is straightforwardly generalizable 
to mixtures since the
functionals for the reference hard--sphere mixtures are known.
First studies of 
Lennard--Jones mixtures \cite{Kah96} confirm the accuracy of the approach.
Besides the computation of the pair structure in fluids, the 
connection to density functional
theory makes FHNC also a versatile tool to study wetting/drying 
phenomena \cite{Oet05}
and effective depletion potentials in dilute
colloidal solutions \cite{Amo01,Oet04,Aya05,Amo07}.

\section{Conclusion}\label{sec6}

In the present work, two coarse-grained models of quadrupolar
fluids were compared with each other (and with both experiment and a 
theory  (IE/DF) 
combining an integral equation approach with 
density functional theory). The aim of our works is to develop 
some understanding for which region of parameters such coarse
grained models are accurate, and also to clarify the applicability
of the analytic theory. While we use experimental data for carbon
dioxide as a prototype example of a quadrupolar fluid for comparison,
our aim is not to provide a chemically realistic modeling description
of this substance or any other material. Recalling that there is 
a lot of interest to use supercritical carbon dioxide as a solvent
and for chemical processing \cite{6,7,8}, and that other quadrupolar 
fluids such as benzene also find widespread applications, there is a need for 
efficient coarse-grained models of such simple molecular fluids. 
(A chemically realistic modeling of systems like polystyrene-carbon
dioxide mixtures would be far beyond reach). Due to the fact that critical 
fluctuations invalidate simple analytic theories extending the 
Van-der-Waals approach (see \cite{56} for a discussion), a simulation
approach as presented here is well-suited to include a sufficiently
accurate description of the critical region.

We model the quadrupolar fluid by spherical point particles carrying a quadrupole
moment (the strength of this quadrupole moment being taken from 
experiment), so that the total interaction between two molecules
is the sum of a Lennard-Jones interaction (Eq.\ \ref{eq1}) and the
quadrupole-quadrupole interaction (Eqs.\ \ref{eq2}, \ref{eq3}).
Possible three-body forces are not at all included explicitly, but
to some extent implicitly, since the Lennard-Jones 
parameters of our effective potential are chosen such that the
actual critical temperature and density of the material (CO$_2$ in
the chosen example) are reproduced. For the sake of computational 
efficency, the potential is truncated at a cutoff distance $r_c$
and shifted to zero there (Eq.\ \ref{eq4}).

As a second model we chose a closely related one, where the angular
dependence of the quadrupolar interaction is averaged perturbatively,
Eq.\ (\ref{eq6}). This isotropic potential can as easily be treated
numerically (sec.\ \ref{IE/DF}) as other simple isotropic pairwise 
potentials.

By construction, the two models have to agree at very high temperatures,
but this is not the region of interest for applications. In the critical 
region, discrepancies of the order of 4\% are typically found, in the case
of carbon dioxide.

However, when we determine the effective Lennard-Jones parameters
for both models such that they reproduce the same critical temperature
and density (namely the critical parameters of carbon dioxide in our
case), we find that both models give a similarly accurate description
of the equation of state over a rather wide region of parameters
(Fig.\ \ref{fig6}, \ref{fig7}, \ref{fig10}). Considering also the 
surface tension (Fig.\ \ref{fig8}), the simpler model with the averaged
interactions is in better agreement with experiments, despite the fact that at high
enough temperatures, relative deviations of the averaged model from
experiment of the order of 1-2\% can be identified (see Fig.\ \ref{fig9}
for example). But even for quantities such as isobars at p=100 bar and 200 bar
(Fig.\ \ref{fig10}), about three times the critical pressure, the experimental
results are very well reproduced over the full density region of 
interest (from 0.1 g/cm$^3$ to 1.0 g/cm$^3$). For such data outside
of the critical region, our IE/DF theory yields a quantitatively 
accurate description without any adjustable parameter whatsoever, 
provided we use the Lennard-Jones parameters obtained from the Monte Carlo
study as an input (fitting the Lennard-Jones parameter to the critical 
parameters of the analytic theory is not appropriate, of course, since the 
latter is inaccurate).

Of course the fact that the simple isotropic model is even 
slightly ''better'' than the more complicated one, as far as the 
comparison with experiment goes, must be attributed to some
lucky cancellation of errors. 
In particular, the temperature dependence of the interface tension
(Fig.\ \ref{fig8}) suggests that the full 
model might itself be too simple for an
optimal description of the fluid, and it may be necessary to include
additional effects like the non-spherical shape of the molecule.
(The model with the angular-dependent
interaction is still far from a full description of chemical reality,
of course: but the comparison presented in \cite{27} shows that 
many very sophisticated atomistic models do not perform better
than the current simple isotropic model either). 
A possible
explanation of this could stay in the fact that the physical
quadrupolar moment used in coarse grained  interactions 
(Eqs.\ \ref{eq4} and \ref{eq6}) could require some effective corrections
related to the truncation of our potentials.
 However, also in view of our need of efficient coarse grained models,  
 the fact that the averaged model is definitively faster than 
the model in which angular degrees of freedom are taken into account,
leads to 
the conclusions of the present work, namely we
 strongly support the use of the averaged 
models \cite{27,29}.

We hence suggest that the present approach using effective
isotropic models for quadrupolar fluids in spite of the 
angular dependence of the interactions in such systems is useful
and we plan to extend it to binary fluid mixtures
as well.
\\
\\
{\bf ACKNOWLEDGEMENTS}\\
B.M.M.\ thanks the BASF AG (Ludwigshafen) for financial support,
while  M.O. was supported by the Deutsche Forschungsgemeinschaft via the
Collaborative Research Centre (Sonderforschungsbereich) SFB-TR6 "Colloids
in External Fields" (project section D6-NWG). 
CPU times was provided by the NIC J\"ulich and the  ZDV Mainz.
Useful and stimulating discussion with F.Heilmann,
L.G.MacDowell, M.M\"uller and H.Weiss are gratefully
acknowledged.

\begin{figure}\caption{\label{fig2} }
\includegraphics[angle=-90,scale=0.45]{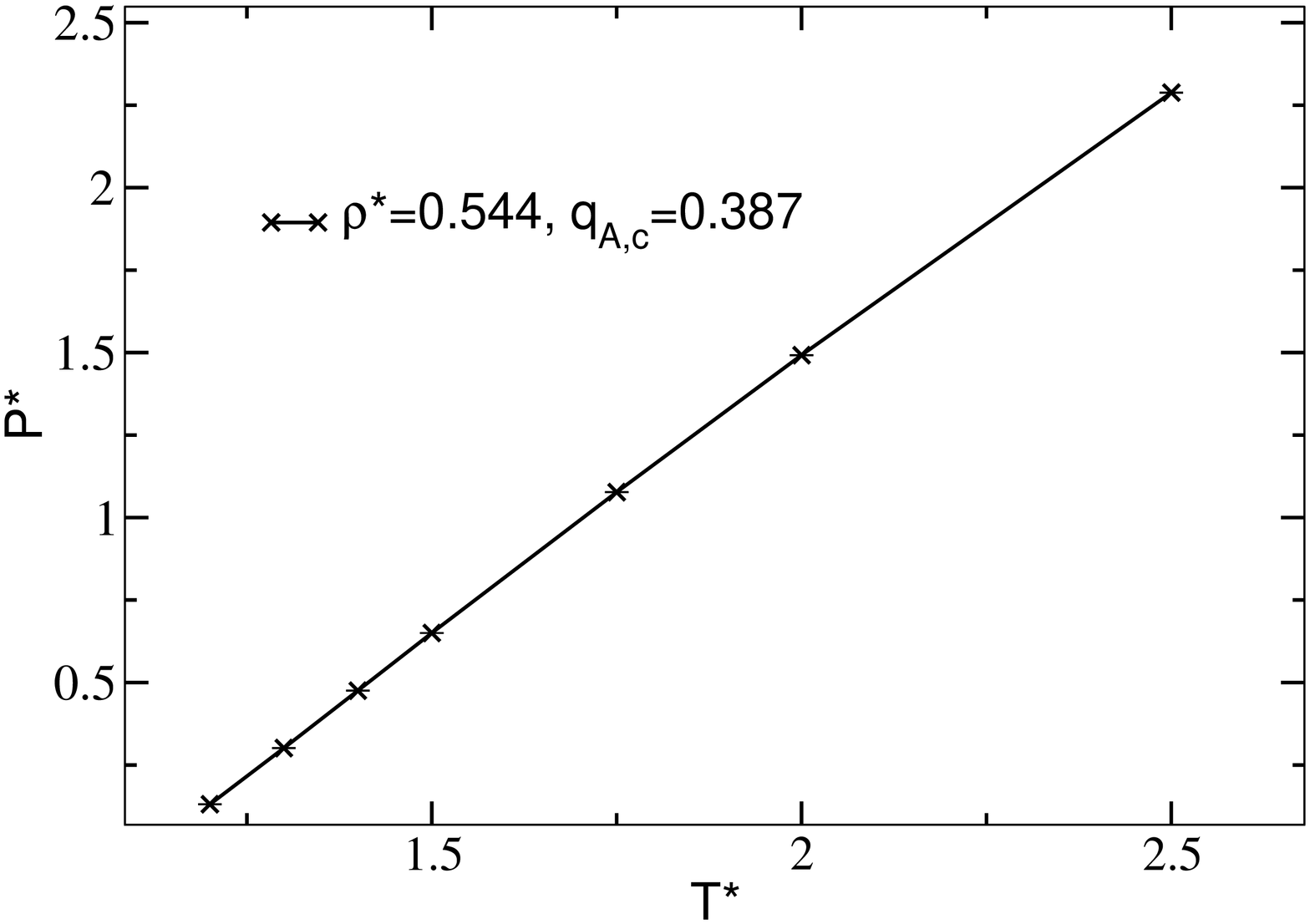}
\includegraphics[angle=-90,scale=0.45]{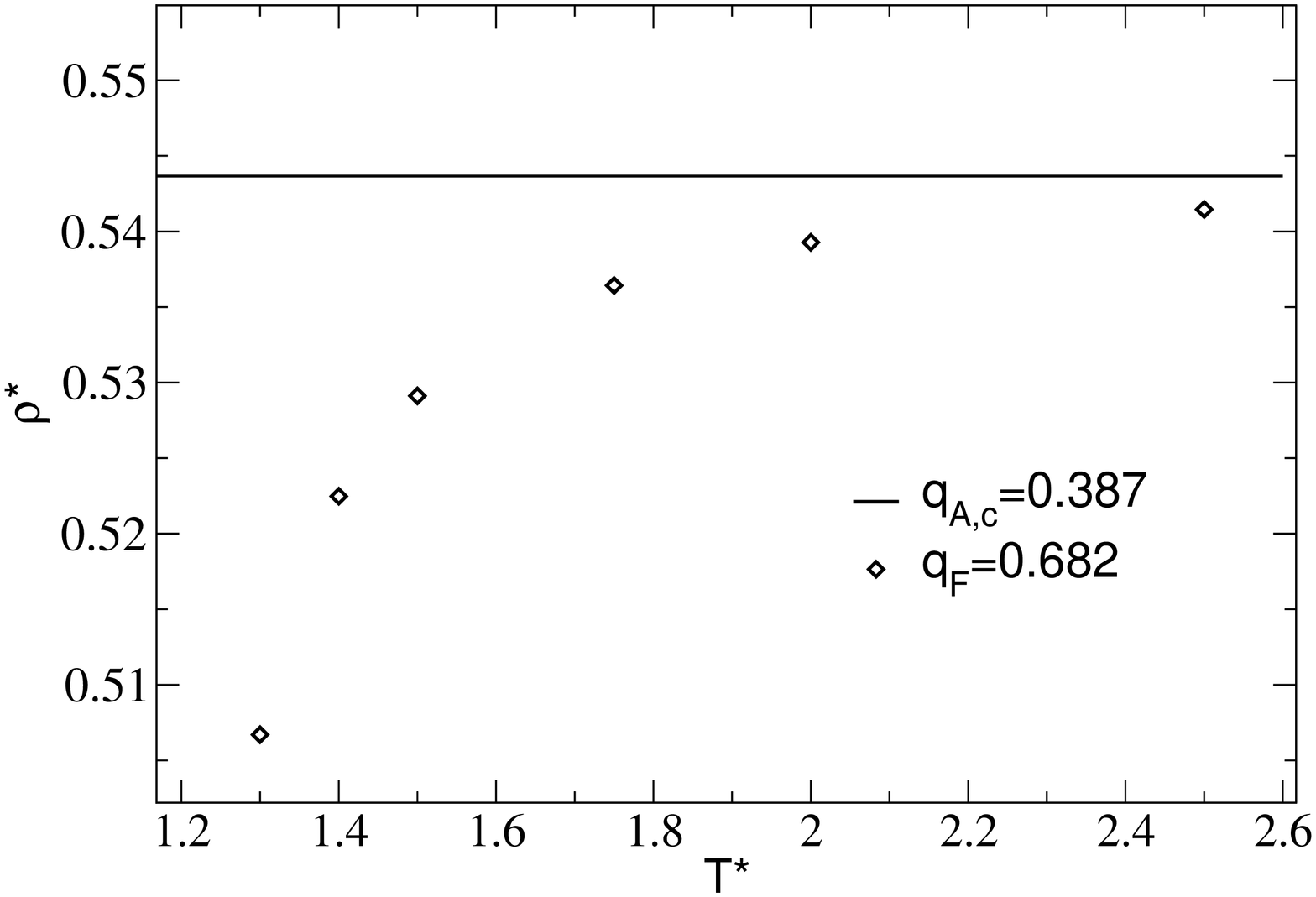}
\end{figure}

\begin{figure}\caption{\label{fig3} }
\includegraphics[angle=-0,scale=0.7]{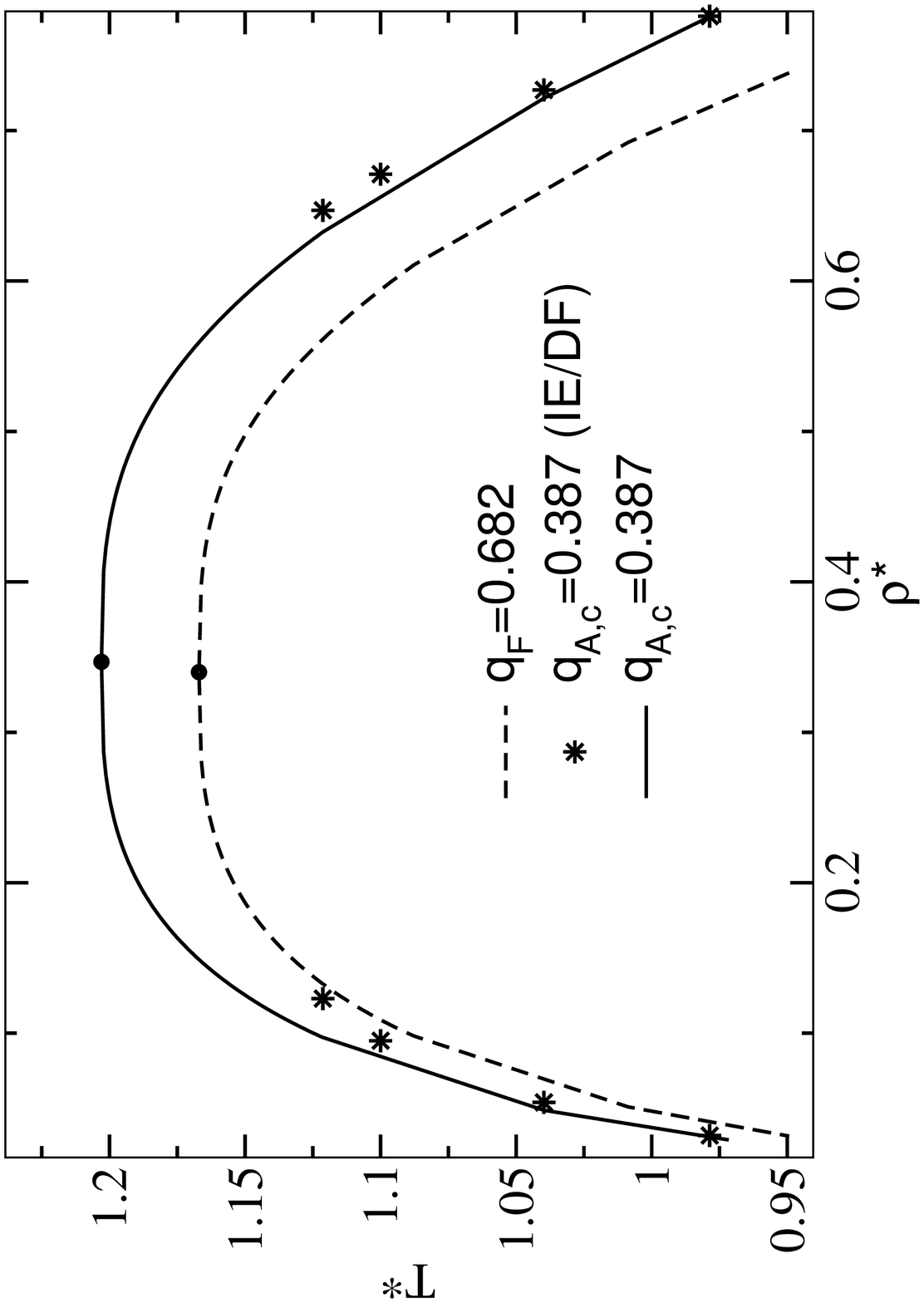}
\end{figure}

\begin{figure}\caption{\label{fig4}}
\includegraphics[angle=-0,scale=0.7]{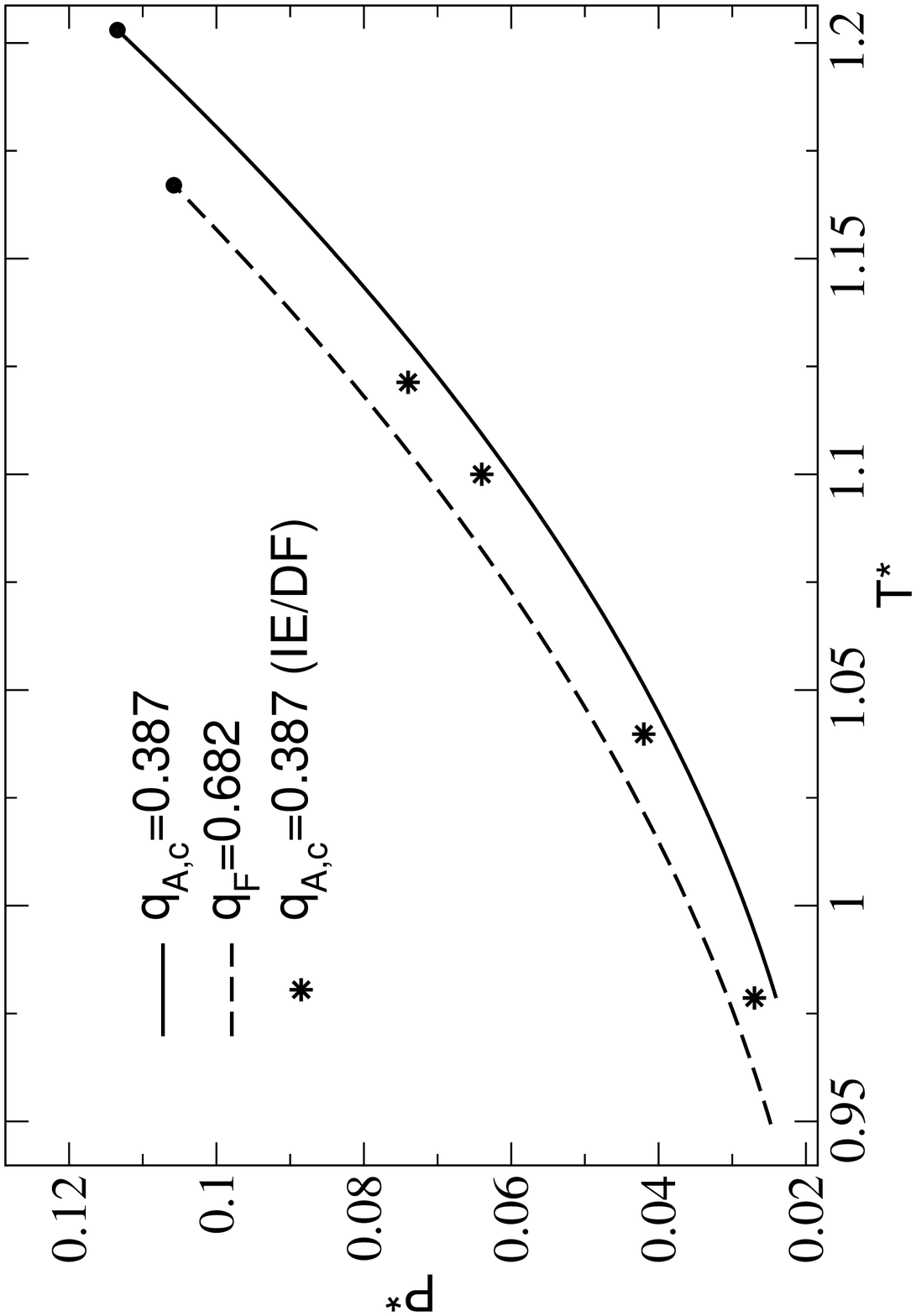}
\end{figure}

\begin{figure}\caption{\label{fig5}}
\includegraphics[angle=-0,scale=0.7]{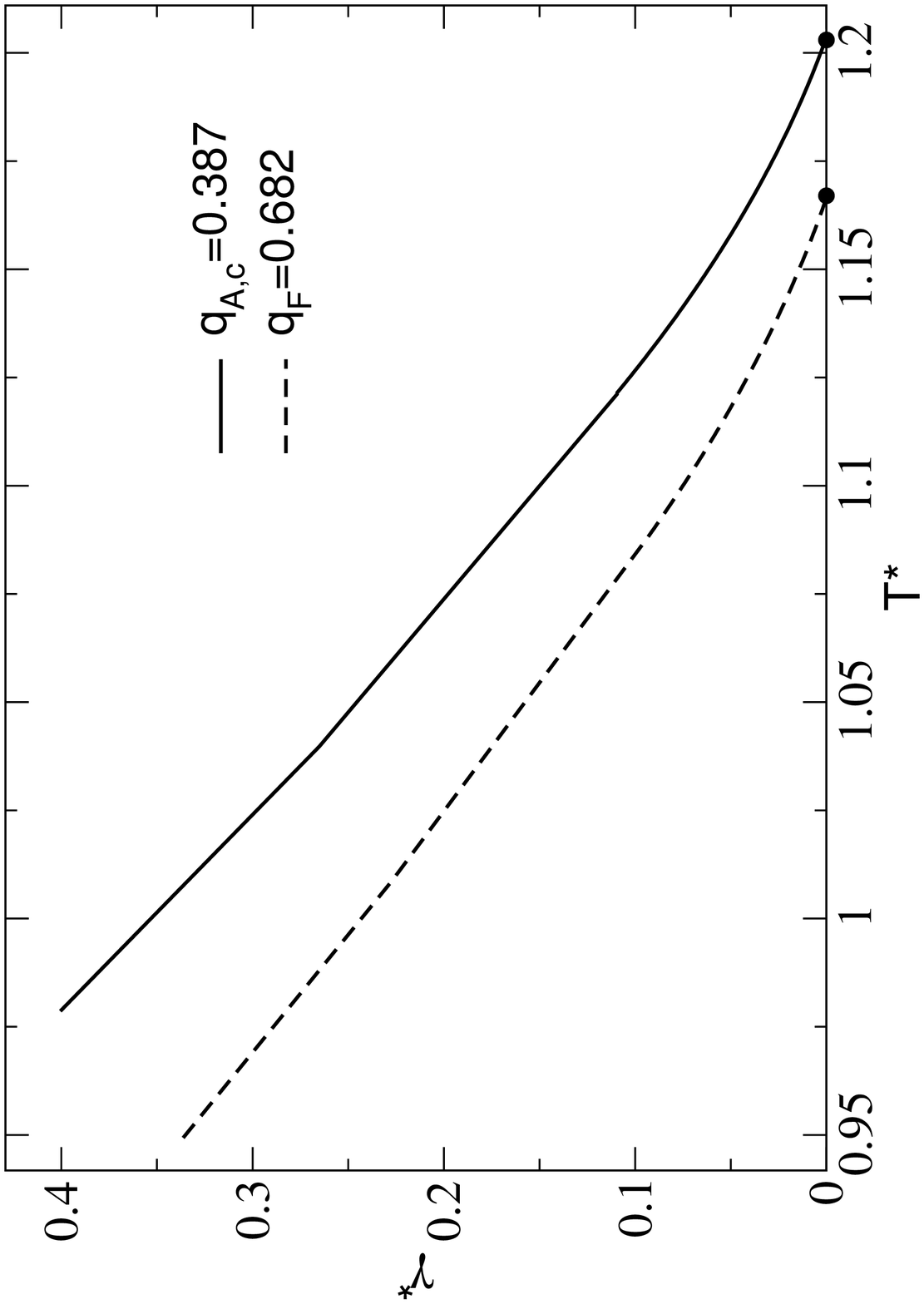}
\end{figure}

\begin{figure}\caption{\label{fig6}}
\includegraphics[angle=-0,scale=0.7]{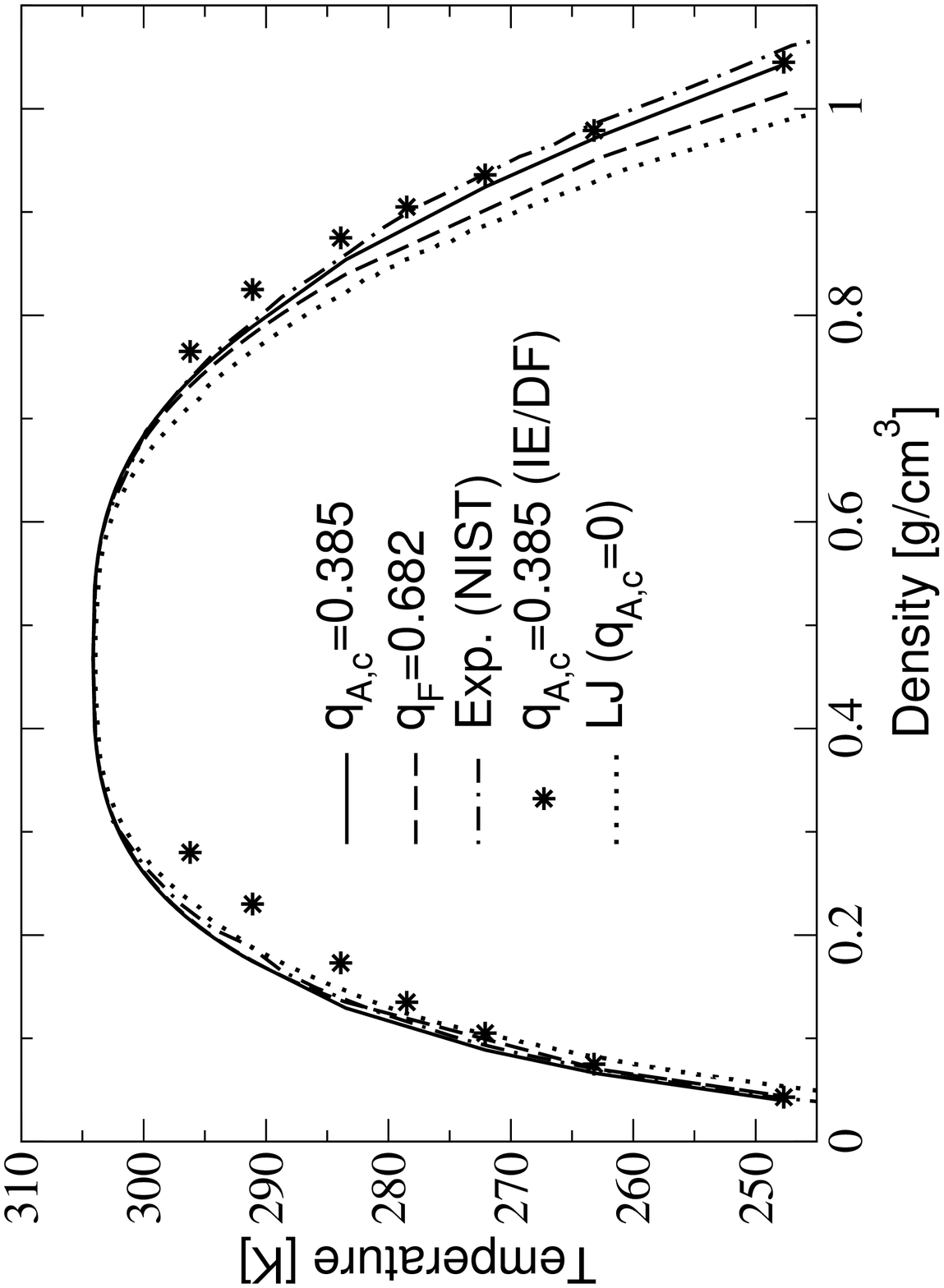}
\end{figure}

\begin{figure}\caption{\label{fig7}}
\includegraphics[angle=-0,scale=0.7]{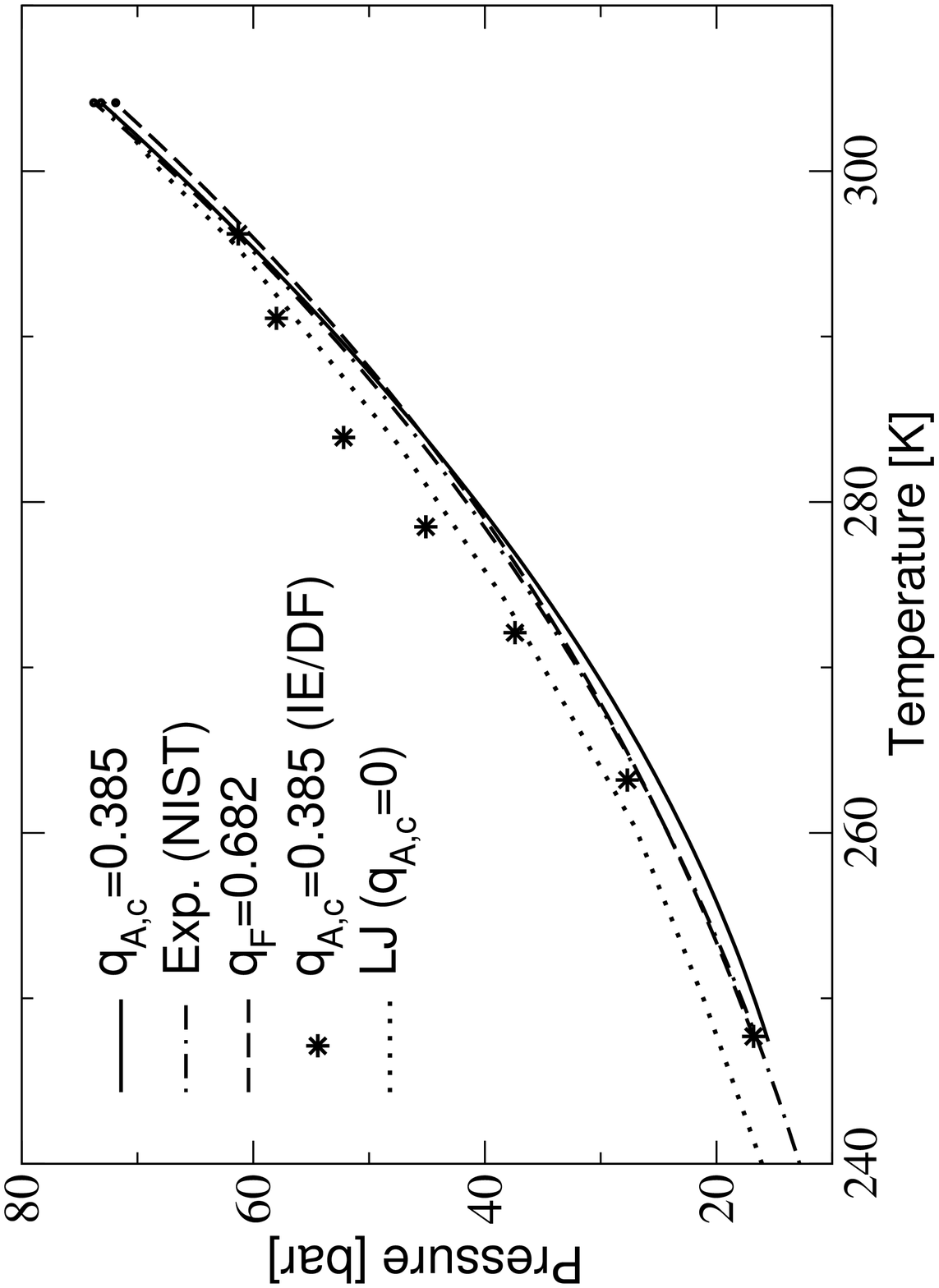}
\end{figure}

\begin{figure}\caption{\label{fig8}}
\includegraphics[angle=-0,scale=0.7]{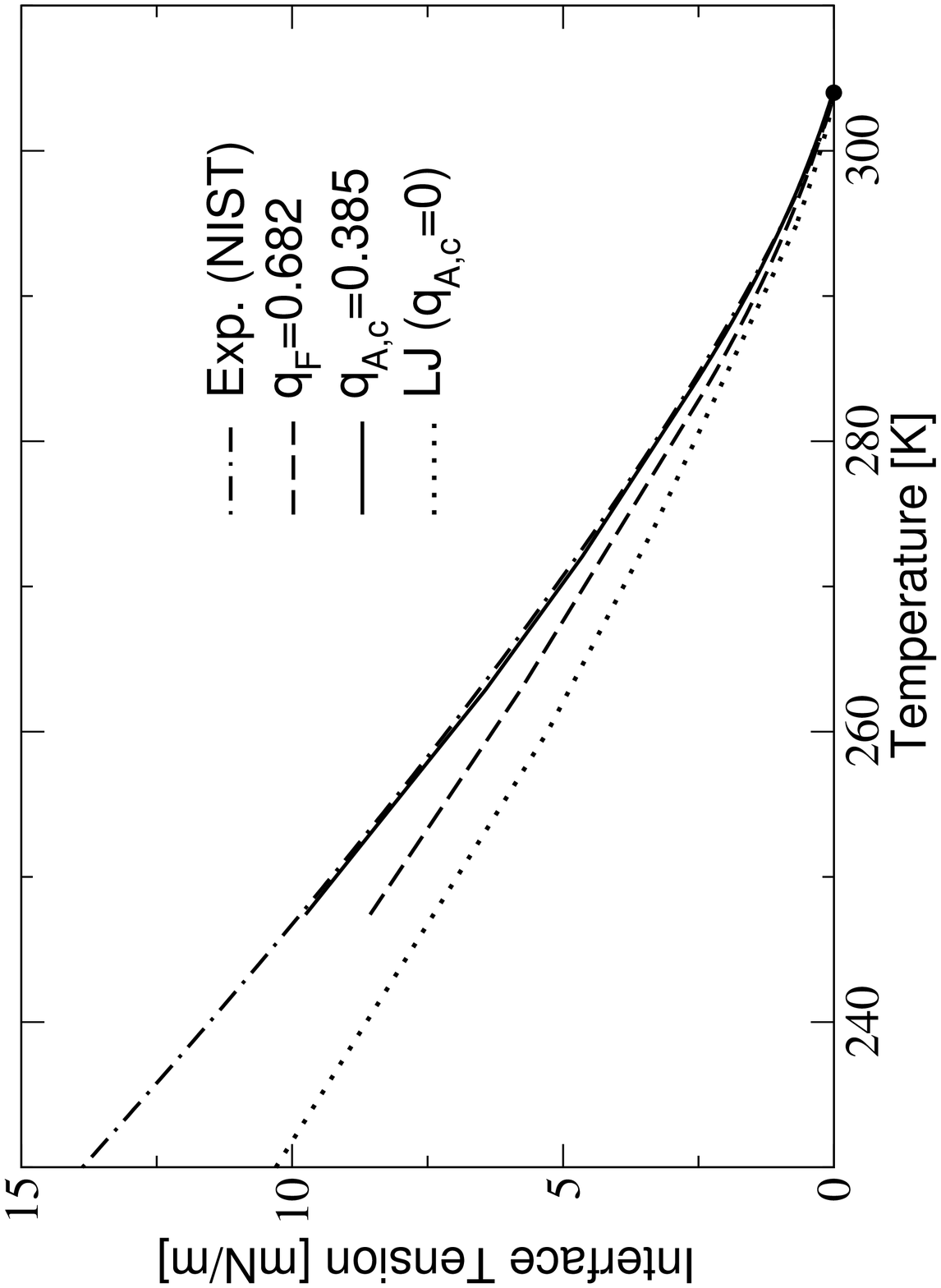}
\end{figure}

\begin{figure}\caption{\label{fig9}}
\includegraphics[angle=-90,scale=0.45]{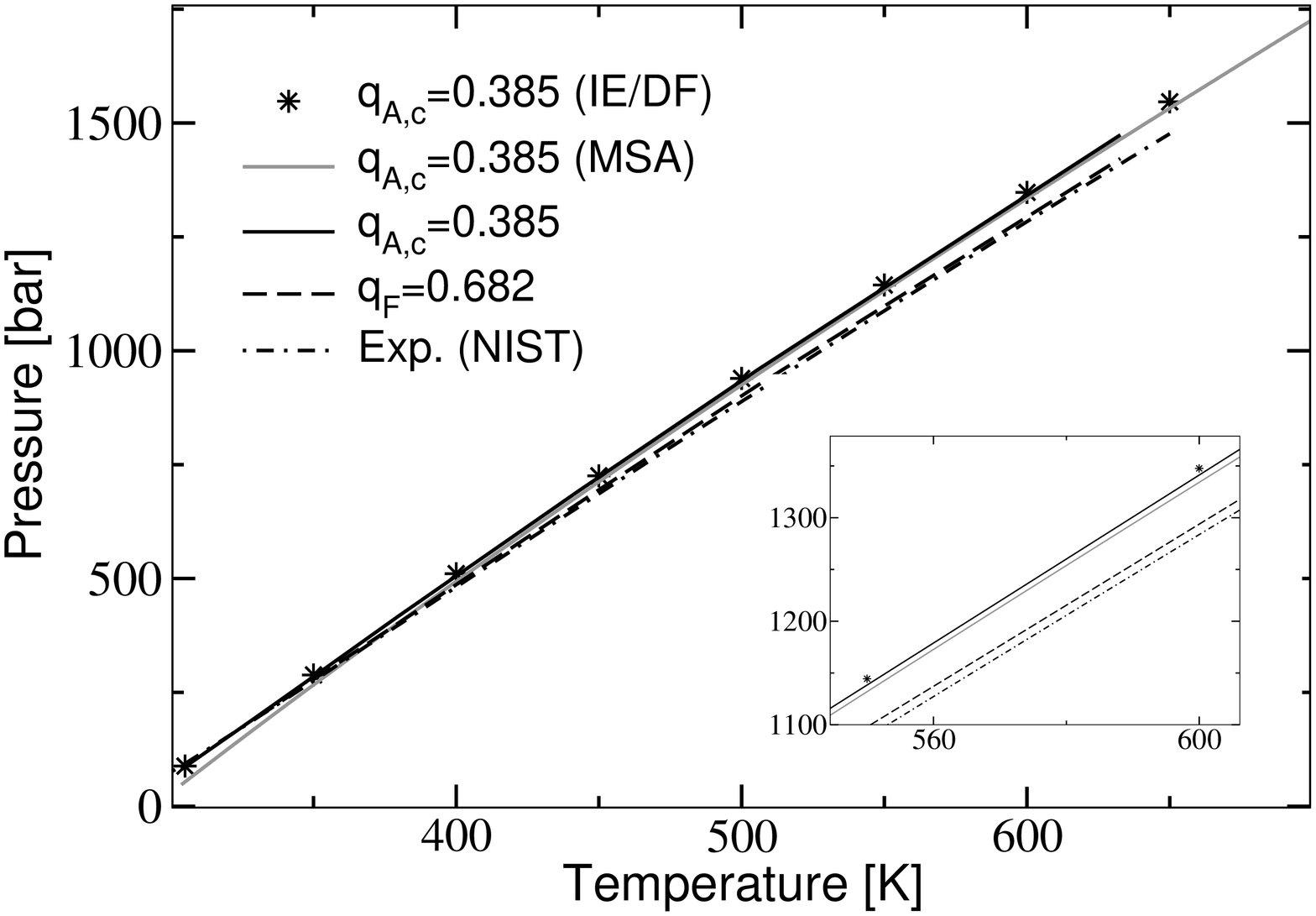}
\includegraphics[angle=-90,scale=0.45]{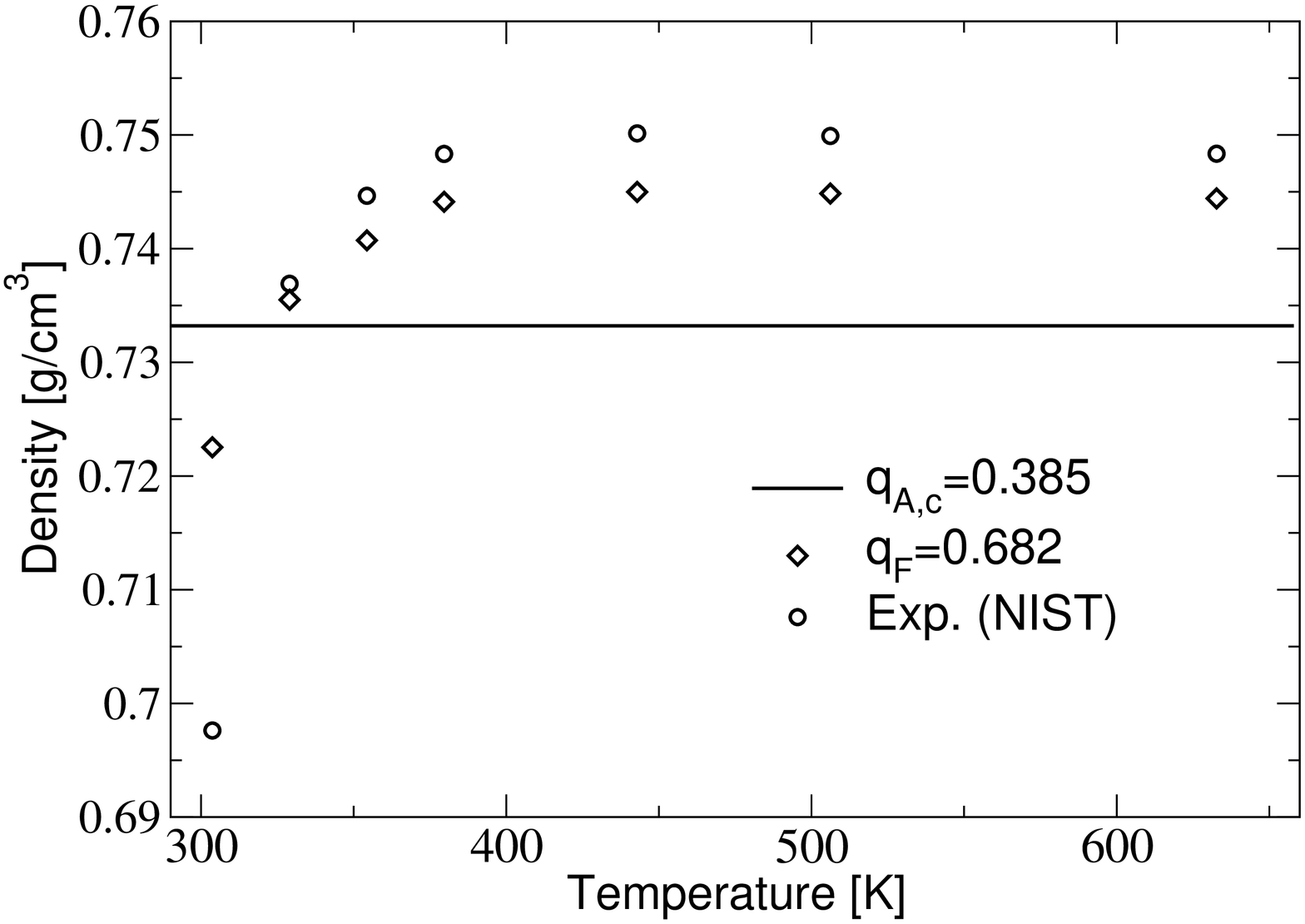}
\end{figure}

\begin{figure}\caption{\label{fig10}}
\includegraphics[angle=-0,scale=0.7]{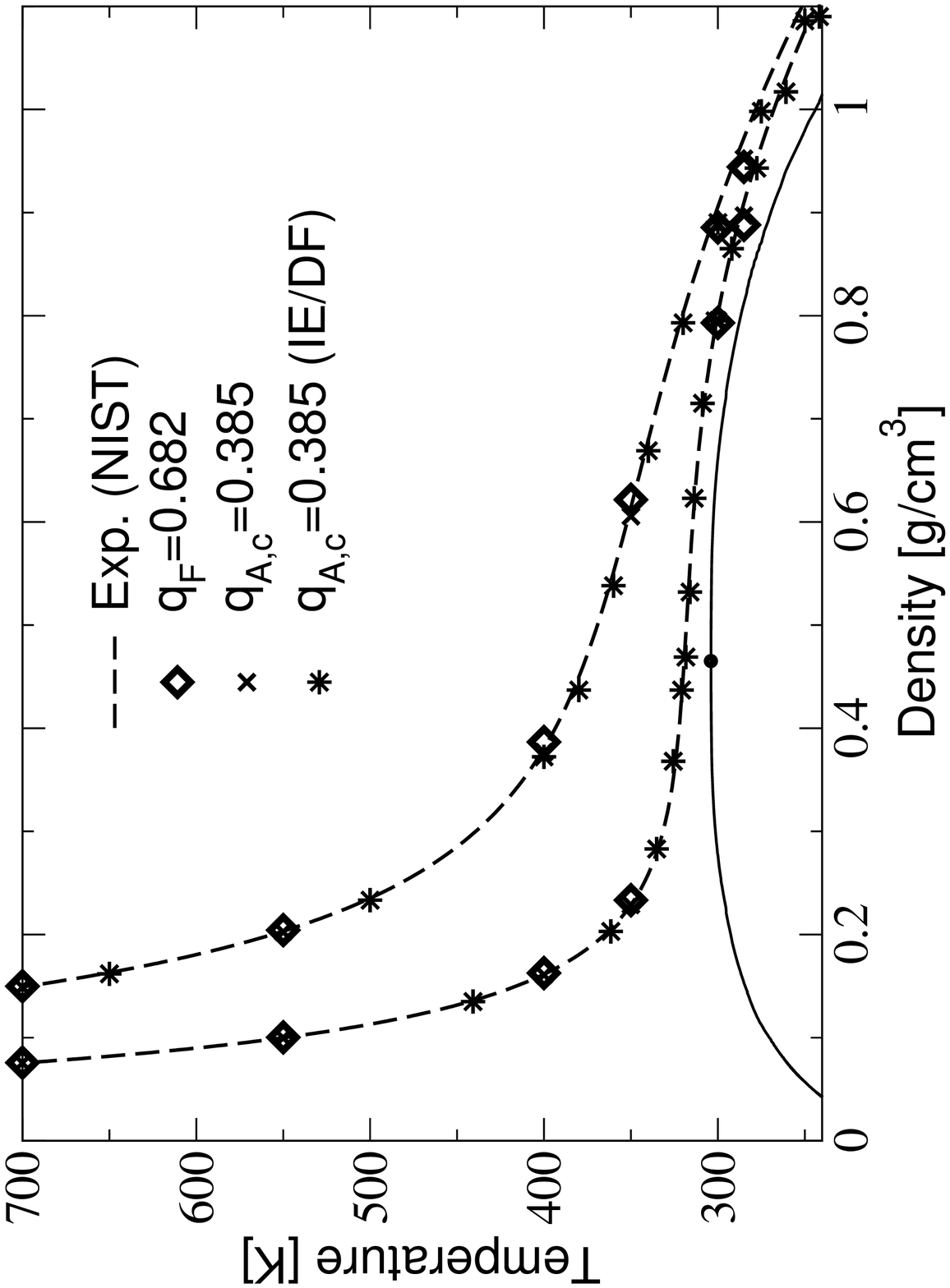}
\end{figure}

\begin{figure}\caption{\label{fig11}}
\includegraphics[angle=-0,scale=0.7]{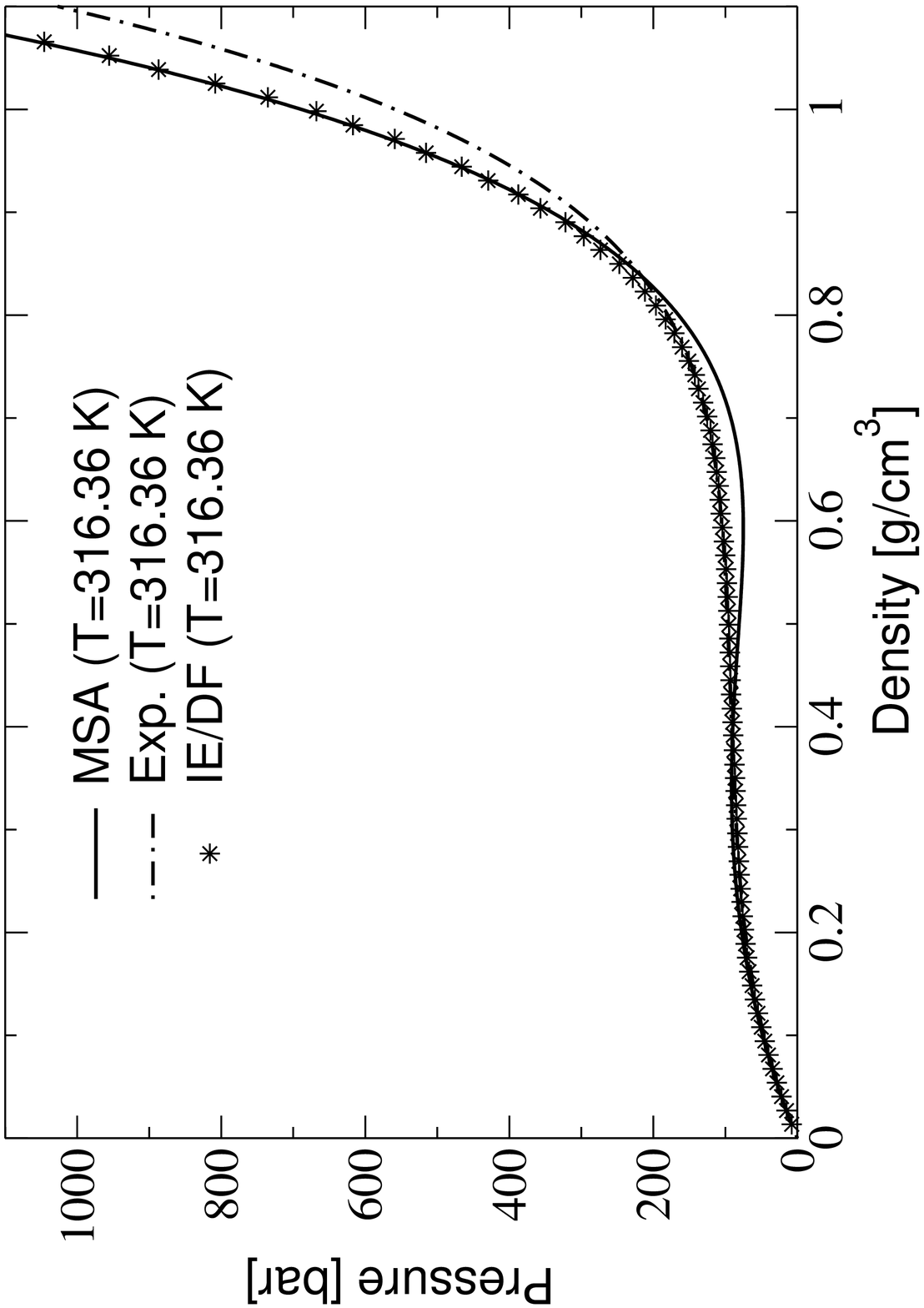}
\end{figure}

\newpage
FIG.\ 1:
Second and fourth order
cumulants plotted vs. $T^*=T/\epsilon$ for $q_F=0.682$ using the
model defined by Eqs.~(\ref{eq3},\ref{eq4}), for four choices of
$L$, namely $L/\sigma = 6.74, 9, 11.3$ and 13.5 (note that the
slope of these curves increases with $L$). The dotted horizontal
lines indicate the theoretical values for the 3d-Ising universality
class \cite{40}.
\\ \\
FIG.\ 2: (a) Reduced pressure $p^*(T)$ 
(in units of the parameters of Eq.\ \ref{eq8}) plotted vs.\
 reduced temperature
$T^*$ for the reduced density $\rho^*=0.544$, 
as obtained from the averaged potential,
Eq.~(\ref{eq6}) (see $\times$). 
(b) Reduced density $\rho^*$ plotted versus reduced
temperature, when one takes the pressure from part (a), as input for an
NpT simulation, using the full potential,
Eqs.~(\ref{eq3}), (\ref{eq4}), with the parameters chosen as given in
Eq.~(\ref{eq8}) (see $\diamond$). Note that the statistical errors are
estimated not to exceed the size of the symbols. 
\\ \\
FIG.\ 3: Vapor-liquid coexistence curve
in the $(T^*,\rho^*)$ plane as predicted by Eq.~(\ref{eq6}) (full line),
 using
the parameters as quoted in Eq.~(\ref{eq8}), and as predicted by
Eqs.~(\ref{eq3}), (\ref{eq4}) (broken line), using the corresponding 
parameters
(Eqs.~\ref{eq7}, \ref{eq8}) implying exact agreement between
both models in the limit $T \rightarrow \infty$. For the 
averaged model, we also report results of
the integral equation/density functional theory described 
in Sec.\ \ref{IE/DF} (see $\ast$). The relative accuracy of the 
curves representing simulation results in this figure and
in the following is estimated to be better than 0.5\%.
\\ \\
FIG.\ 4: Vapor-liquid coexistence curve
in the $(p^*,T^*)$ plane, for the same choices as in
Fig.~\ref{fig3}. For the 
averaged model, we report also results of
the integral equation/density functional theory described 
in Sec.\ \ref{IE/DF} (see $\ast$).
\\ \\
FIG.\ 5: Interfacial tension plotted
vs. $T^*$, for the two models as specified in Fig.~\ref{fig3}.
\\ \\
FIG.\ 6:
Same as Fig.~\ref{fig3}, but
choosing the parameters of Eq.\ (\ref{eq11}) for the model with full
quadrupolar interaction (broken line) and of Eq.\ (\ref{eq12}) 
for the model with
the averaged interaction (full line). 
Experimental data from Ref.~\cite{34}
are included (broken-dotted line).
With respect to Fig.~\ref{fig3}, 
critical temperature and density for both models now coincide 
with the experimental values.
We also include the LJ predictions of Ref.\ \cite{28}
(dotted line).
We notice that the spherical and averaged model in this
reparametrized plot produces coexistence densities that are 
in good agreement.
 Indeed differences   are  comparable to
 the two models used to describe CO$_2$ in our previous
paper \cite{27} (i.e.\ $q_\mathrm{A,c}=0.387$ and $q_\mathrm{A,c}=0.470$) 
and with discrepancies of the Lennard Jones model in predicting
the phase diagram for noble gases (see Fig.\ 11 of Ref.~\cite{27}).
Finally we note a better agreement of the averaged model with 
experimental results. For the 
averaged model, we also report results of
the integral equation/density functional theory described 
in Sec.\ \ref{IE/DF} (see $\ast$).
\\ \\
FIG.\ 7:
Coexistence pressure. We report the results for the 
full model with simulation parameters given in Eq.\ (\ref{eq11}) 
(broken line), the results
for the averaged model with simulation parameters reported in Eq.\
(\ref{eq12}) (full line)
and the experimental results  \cite{34} (broken-dotted line).
We also include  the LJ predictions of Ref.\ \cite{28}
(dotted line).
We observe that at low temperatures the full model gives better
results with respect to the averaged model. This is due to the fact
that for high densities the orientational part of the quadrupolar
interaction becomes more important. On the other hand, near
the critical point the averaged model performs better than the
full model. For the 
averaged model, we report also results of
the integral equation/density functional theory described 
in Sec.\ \ref{IE/DF} (see $\ast$).
\\ \\
FIG.\ 8:
Prediction of interface tension. We report the results for the 
full model with simulation parameters given in Eq.\ (\ref{eq11}) (broken line), the results
for the averaged model with simulation parameters reported in Eq.\ (\ref{eq12})
(full line)
and the experimental results \cite{34} (broken-dotted line).
We also include  the LJ predictions of Ref.\ \cite{28}
(dotted line).
 The averaged model is in perfect
agreement with experimental results. 
\\ \\
FIG.\ 9:
(a) Supercritical isochore ($\rho$=0.733 g/cm$^3$)  for the averaged
model with parameters given in Eq.\ (\ref{eq11}) (full line) and for 
the full model with
parameters given in Eq.\ (\ref{eq12}) (broken line). 
Experimental results are also included 
\cite{34} (broken-dotted line). We observe  a very nice agreement between 
the experimental
results and the full potential. 
The small discrepancy between
the averaged model and the full model at high temperatures can be
understood in the light of the results of Fig.~\ref{fig2}. 
Indeed, due to the
fact that the use of the same $\epsilon$  and $\sigma$ 
for the averaged and full
models (see Eq.~\ref{eq8}) produces the same equilibrium
 states at high temperature (see 
Fig.~\ref{fig2}), the
new choice of parameters Eqs.\ (\ref{eq11}) (\ref{eq12})
 produces a systematic discrepancy
at high temperature. 
 For the averaged model, we also report results of
the integral equation/density functional theory described 
in Sec.\ \ref{IE/DF} ($\ast$) and the results of the perturbative
MSA theory described in \cite{27} (gray line). The inserted picture 
shows as both the theory are in very nice agreement with the MC 
results.
\\
 (b). Similarly as in Fig.\ \ref{fig2}(b) we report the 
prediction  for the densities 
for the full model (diamond) obtained in an NpT simulation
taking as  input the pressures plotted in Fig.\ \ref{fig9}(a) for the averaged
model (full line). 
The resulting densities 
agree well with experiments \cite{34}. The
small deviation at high temperature (between the two models)
can be explained by considerations discussed for Figs.\ \ref{fig2}
and \ref{fig9}(a).
\\ \\
FIG.\ 10:
Supercritical isobars (p=100 and  p=200 bar). We report the results for the 
full model with simulation parameters given in Eq.\  (\ref{eq11}), the results
for the averaged model with simulation parameters reported in Eq.\ (\ref{eq12})
and the experimental results \cite{34}. The agreement 
is very good. For a comparison with two other averaged models
we refer to Fig.\ 8 of our previous paper \cite{27}.
 For the averaged model, we report also results of
the integral equation/density functional theory described 
in Sec.\ \ref{IE/DF}.
\\ \\
FIG.\ 11:
Comparison between the Integral Equation/Density Functional
(IE/DFT) theory ($\ast$) 
and an equation of state in the perturbative Mean Spherical Approximation
(MSA) described in \cite{27,MMVB00,MVMB02} (full line) for
the supercritical isotherm T=316.36 K. 
IE/DFT,  if compared to experiments \cite{34}, performs better  
for intermediate densities 0.3 g/cm$^3$ $<\rho< $0.8 g/cm$^3$. However
for $\rho>0.8$ g/cm$^3$ the two theory predict almost the same
 equilibria states.
\newpage
\clearpage

\newpage
\clearpage

\newpage
\clearpage

\newpage
\clearpage

\newpage
\clearpage

\newpage
\clearpage

\newpage
\clearpage

\newpage
\clearpage

\newpage
\clearpage

\newpage
\clearpage

\newpage
\clearpage

\end{document}